\documentclass[]{aa}

\usepackage{rotating}
\usepackage{graphicx}
\usepackage{sidecap}
\usepackage{subcaption}
\usepackage{url}
\usepackage[colorlinks=true]{hyperref}
\usepackage{soul}
\hypersetup{allcolors=blue}
\usepackage[normalem]{ulem}

\usepackage[kerning=true, tracking=true]{microtype}

\usepackage{lipsum}
\usepackage{txfonts}
\usepackage[]{xcolor}
\usepackage{float}

\bibpunct{(}{)}{;}{a}{}{,}
\DeclareUnicodeCharacter{00A0}{~}

\newcommand{\kms}{\,km\,s$^{-1}$}

\newcommand{\Halpha}{H$\alpha$}

\newcommand{\dyn}{\,dyn\,cm$^{-2}$}
\usepackage{cancel}
\usepackage{color}

\definecolor{deepmagenta}{rgb}{0.8, 0.0, 0.8}
\definecolor{green}{rgb}{0.1, 0.7, 0.1}

\usepackage{algorithm}
\usepackage{algpseudocode}

\begin{document}






\title{Thermodynamic and magnetic evolution of an eruptive C-class solar flare observed with SST/TRIPPEL-SP}

\titlerunning{Thermodynamics and magnetic evolution of an eruptive C-class~flare}
\author{C. J. D\'iaz Baso
        \inst{1,2}
        \and
        J. de la Cruz Rodr\'iguez
        \inst{3}
        \and
        H.-P. Doerr
        \inst{4,5}
        \and
        M. van Noort
        \inst{4}
        \and
        A. Prasad
        \inst{6,1,2}
        \and\\
        A. Feller
        \inst{4}
        \and
        D. Kiselman
        \inst{3}
        }
        \institute{
        Institute of Theoretical Astrophysics,
        University of Oslo, %
        P.O. Box 1029 Blindern, N-0315 Oslo, Norway
        \and
        Rosseland Centre for Solar Physics,
        University of Oslo, %
        P.O. Box 1029 Blindern, N-0315 Oslo, Norway
        \and
        Institute for Solar Physics, Dept. of Astronomy, Stockholm University, AlbaNova University Centre, SE-10691 Stockholm, Sweden
        \and
        Max Planck Institute for Solar System Research, Justus-von-Liebig-Weg 3, 37077 G\"ottingen, Germany
        \and
        Th\"uringer Landessternwarte, Sternwarte 5, 07778 Tautenburg, Germany
        \and
        Statkraft AS, Lysaker, Norway
        \\
        \email{carlos.diaz@astro.uio.no}
        }


\authorrunning{D\'iaz Baso et al.}
\abstract
{
Solar flares are complex phenomena driven by the release of magnetic energy, but a large energy reservoir is not sufficient to determine their eruptive potential; the magnetic topology and plasma dynamics play a key role. We investigate the thermodynamic and magnetic properties of the solar atmosphere during the rise, peak, and decay phases of a C5.1-class flare and filament eruption in active region NOAA 12561 on 2016 July 7, to understand the origin and atmospheric response of this event. High spatial and spectral resolution spectropolarimetric observations of the chromospheric $\ion{Ca}{ii}$ 8542\,\AA\ line and nearby photospheric lines were obtained with the TRIPPEL-SP spectropolarimeter at the Swedish 1-m Solar Telescope. Using non-local thermodynamic equilibrium (NLTE) inversions and non-force-free field (NFFF) magnetic extrapolations, we followed the event's evolution from its precursor to its decay. Before the flare, our analysis reveals a complex, sheared magnetic topology with a high free energy content ($\sim2\times10^{30}$ erg). In this precursor phase, we detected persistent, localized heating (temperature increase of $\sim$2000 K) with strong downflows ($\sim$10-20 km s$^{-1}$) deep in the atmosphere. This heating was co-spatial with a bald-patch region, suggesting that low-altitude magnetic reconnection could destabilize the filament of the region. The flare's rise phase was marked by the filament's eruption, with a total speed larger than $\sim$70~\kms, when combining inversions and plane-of-sky motions. Following the eruption, the free energy decreased by $\sim$30\% as post-flare loops formed, connecting the flare ribbons and channeling the released energy into the lower atmosphere. The flare ribbons exhibited significant heating to $\sim$8500 K and downflows up to $\sim$10~\kms, consistent with energy deposition along reconnected loops.
}


\maketitle



\section{Introduction}\label{sec:intro}

Solar flares are sudden and intense energy release phenomena occurring in the solar atmosphere. They manifest across the entire electromagnetic spectrum, with an impulsive phase that lasts from seconds to minutes, and involve rapid plasma heating, particle acceleration, and significant restructuring of the magnetic field \citep[e.g.,][]{Fletcher2011SSRv}. The energy released is believed to originate from the conversion of magnetic energy stored in stressed, non-potential magnetic field configurations. Understanding the mechanisms of this energy conversion and release is a central goal in solar physics.

While the “storage and release” model of \citet{Rosner1978} suggests that the flare magnitude should scale with the magnetic energy reservoir, recent statistical studies \citep[e.g.,][]{Karimov2024} find little correlation between estimated stored energy and flare activity as measured by GOES X-ray flux. This highlights the importance of flare triggering mechanisms beyond just energy storage. Triggers may involve large-scale MHD instabilities, such as flux-rope eruptions \citep[e.g.,][]{Torok2005ApJ}, or smaller-scale reconnection events in the lower atmosphere \citep[e.g.,][]{Moore2001ApJ} that destabilize the overlying field. Observational signatures like precursor brightenings or specific flow patterns often serve as clues to these initiation processes \citep[e.g.,][]{Sterling2015Natur,Jeffrey2018Sci,Chitta2020,Panos2023A&A}. In particular, magnetic configurations known as "bald patches" (BPs)—locations where magnetic field lines are tangent to the photosphere and dip before arching upward—have been identified as favorable sites for low-altitude reconnection that may lead to chromospheric heating and potentially trigger larger flares \citep{Aulanier2000,2017RAA....17...93J}.

Given that the magnetic field not only stores the energy but also governs the processes that initiate and drive flares \citep[e.g.,][]{Priest2002AARv}, a comprehensive understanding requires detailed knowledge of the magnetic and plasma conditions throughout the solar atmosphere, from the photosphere to the corona. Spectropolarimetry, particularly using spectral lines formed at different atmospheric heights, provides the necessary observables to infer the stratification of temperature, velocity, and magnetic field vector \citep[e.g.,][]{delToroIniesta2016}.

A significant fraction of the total flare energy is radiated away in the chromosphere \citep[e.g.,][]{Milligan2014ApJ}, making it a critical layer for studying energy deposition and dissipation. This layer acts as a crucial interface where energy transported from the corona (e.g., via non-thermal particles or thermal conduction) is deposited and dissipated, leading to intense heating and dynamic responses. Strong chromospheric lines—such as the \ion{Ca}{ii} H \& K lines, \Halpha, \ion{Ca}{ii} 8542\,\AA, and the \ion{He}{i} 10830\,\AA\ triplet—are powerful diagnostics of chromospheric conditions, and have long been used to constrain semi-empirical flare models \citep[e.g.,][]{Machado1980ASSL,Libbrecht2019,Yadav2022_RadLoss}

Among these, the \ion{Ca}{ii} 8542\,\AA\ line has proven particularly valuable for probing the magnetic and thermodynamic structure of the chromosphere \citep[e.g.,][]{Cauzzi2008A&A,DiazBaso2021,Morosin2022A&A,Vissers2022A&A_ARmag}. Its high sensitivity to both temperature and magnetic fields via the Zeeman effect makes it ideal for flare studies across a wide range of magnitudes \citep[e.g.,][]{Kleint2015ApJ,Ferrente2023}. Using non-local thermodynamic equilibrium (NLTE) inversions, several studies have revealed strong temperature enhancements, plasma flows associated with chromospheric condensation and evaporation, and significant changes in the magnetic field during flares \citep[e.g.,][]{Kuridze2017ApJ, Vissers2020_flare, Yadav2021_Cflare, Kuckein2025A&A, Yadav2025_Ribbons}.







Some of the previously cited studies have offered different interpretations of the chromospheric magnetic field changes \citep[e.g.,][]{Kleint2017ApJ,Yadav2021_Cflare} and flow dynamics \citep[e.g.,][]{Ferrente2024,Kuckein2025A&A} during the respective analyzed flares, but it remains unclear whether these differences arise from the energy released, the thermodynamic–magnetic environment, or the analysis techniques employed. 
Therefore, the aim of the present work is to carry out a comprehensive analysis of a C5.1-class flare and associated filament eruption. Our specific objectives are to: (i) search for precursor activity in the lower atmosphere; (ii) characterize the chromospheric dynamics associated with the filament eruption; (iii) quantify the thermodynamic and magnetic response of the chromosphere; (iv) track changes in the 3D magnetic topology and free energy content; and (v) synthesize these results into a coherent scenario.


\section{Observations}\label{sec:observation}

\subsection{Target and observational setup}\label{sec:setup}

The data used in this work have been recorded with a temporary, highly experimental polarimetric add-on for the TRI-Port Polarimetric Echelle-Littrow spectrograph \citep[TRIPPEL;][]{Kiselman2011A&A} at the Swedish 1-m Solar Telescope \citep[SST;][]{Scharmer2003}, dubbed TRIPPEL-SP. The additions included a polarimetric modulator based on ferroelectric liquid crystals (FLCs) borrowed from the Fast Solar Polarimeter \citep[FSP;][]{Iglesias2016A&A}, and four fast CMOS cameras (dual-beam spectrum camera, SPC; slit-jaw camera with two phase-diverse channels, SJC).

TRIPPEL-SP has been developed to combine the basic idea of the FSP (faster-than-seeing polarimetric modulation with individual short-time exposures) with the spectrum restoration scheme of \cite{vanNoort2017A&A...608A..76V}. To this end, the SPC sensors were clipped to $2048\times 2048$ pixels to achieve a frame rate of up to 400\,Hz while still allowing a field of view (FOV) of $\approx 60"\times 33\AA$ at 8540\,\AA. The SJCs have been running strictly synchronized (frame rate and exposure time) to the SPCs, with a band-pass filter identical to the SPCs, to allow the retrieval of the field-dependent point spread function (PSF) for each individual spectrum exposure. For scanning, the correlation tracker of the SST was programmed to slowly move the solar image across the spectrograph slit. This continuous-slew scanning facilitates the image restoration and avoids the unwanted dead-time of the classical stepped raster scanning schemes \citep[for details see][]{Saranathan2021A&A}. 

Our target flare SOL2016-07-07T07:49 
developed in the active region NOAA 12561, which emerged on 2016 July 6. We recorded 5~scans on July 7, where each scan took approximately 12-14 minutes, between 07:36 UT and 08:51 UT. The center of the scan was located near (X, Y) = (525\arcsec, 304\arcsec) heliographic coordinates at the beginning of the observations, corresponding to a heliocentric angle $\theta \approx 40^\circ$ ($\mu = \cos\theta \approx 0.77$). Unfortunately, this event was not observed by RHESSI, Hinode, IRIS, or STEREO-A, limiting coronal and transition region diagnostics.

TRIPPEL-SP recorded full Stokes vector ($I, Q, U, V$) spectropolarimetric data in the spectral range from 8524.7\,\AA\ to 8557.6\,\AA\ ($\sim$$32.9$\AA), with an initial spectral sampling of approximately 34.4\,m\AA\,pix$^{-1}$. This spectral window (see Fig.~\ref{fig:im_qs_profile}) includes the strong chromospheric \ion{Ca}{ii} 8542\,\AA\ line, along with several magnetically sensitive photospheric lines, primarily of \ion{Fe}{i} (e.g., 8526.67\,\AA, 8538.01\,\AA) and \ion{Si}{i} (e.g., 8536.57\,\AA, 8556.78\,\AA). In this spectral window we reduced the frame rate of TRIPPEL-SP to 100\,Hz to maintain a photon-noise limited exposure level. The much reduced sensitivity in this particular wavelength is partially caused by the low efficiency of the grating but mostly by the poor quantum efficiency of the detectors of less than 20\%.

\subsection{Data reduction}

The raw spectrograph data were processed using a standard reduction procedure. This included bad-pixel correction, dark current subtraction, flat-field correction, and precise wavelength calibration. The spectra were then restored using the PSFs inferred from the restoration of broadband filter images of the SJCs following the procedure of \cite{vanNoort2017A&A...608A..76V}. This resulted in a significant increase in spatial resolution, contrast and overall homogeneity of the data. After restoration, the modulated spectra were then demodulated to obtain the Stokes $I, Q, U, V$ parameters for each spectral and spatial point. Residual instrumental polarization and crosstalk effects were further corrected using standard techniques \citep[e.g.,][]{2002A&A...381..668S}. One of the SPCs exhibited an excessive noise pattern, most likely caused by a problem with the camera's temperature stabilization. This issue prevented us from successfully merging the two polarimetric beams so that the data used in this work is effectively single-beam only. This is not an issue in terms of polarimetric cross-talk as image restoration has been shown to be very efficient in reducing seeing-induced cross-talk \citep[][]{2022A&A...668A.151V}.

The wavelength and absolute intensity calibrations were performed by comparing the average spectra with the Fourier Transform Spectrometer \citep[FTS;][]{Neckel1984} atlas taking into account the observed heliocentric angle\footnote{These routines are available in the Python repository\\ \url{https://github.com/ISP-SST/ISPy}}. This process included an optimization procedure which degrades the FTS atlas with a Gaussian spectral profile of different widths. The FWHM of the Gaussian that best reproduces the observations allowed us to estimate the spectral resolution to be between 120 $-$ 150\,m\AA. To improve the signal-to-noise ratio (S/N), we spectrally binned the data by a factor of 2, resulting in a final spectral sampling of $\sim$68\,m\AA\,pix$^{-1}$. The noise level in Stokes $Q, U, V$, estimated from the standard deviation in continuum regions far from the spectral lines, was approximately $(7, 10, 8) \times 10^{-3}$ in units of the average continuum intensity ($I_c$). 
These values refer to restored data, making them not directly comparable to noise levels in raw or seeing-limited instruments \citep[see e.g.][for a discussion of noise in restored S/P data]{2022A&A...668A.151V}.

\section{Overview of the event}\label{sec:overview}

\subsection{GOES context}\label{sec:GOES}

\begin{figure}[t]
\centering
\includegraphics[width=\linewidth]{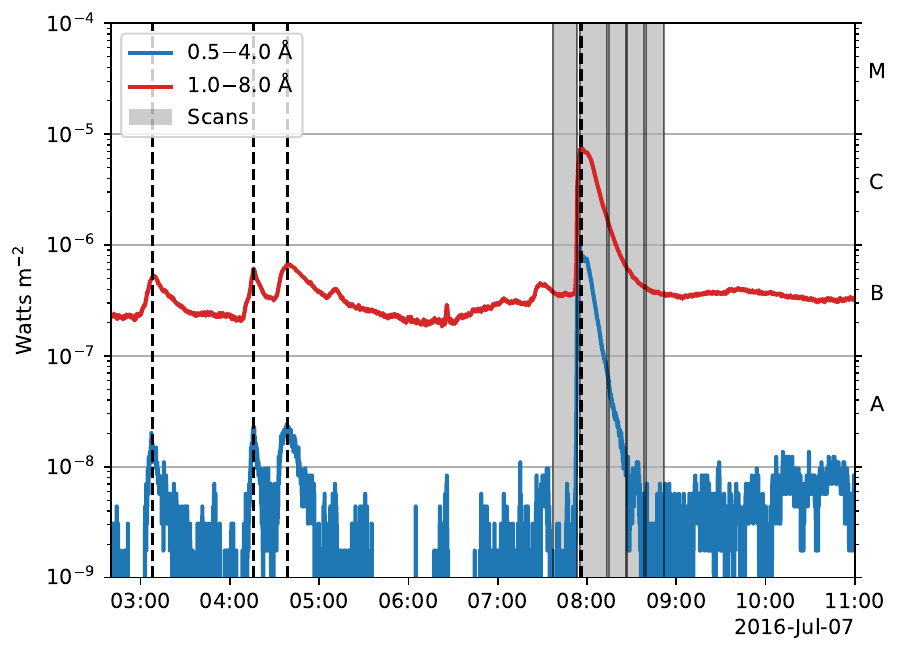}
\caption{Time evolution of GOES X-ray flux in the two channels (1$-$8\AA\ and 0.5$-$4\AA) on 2016 July 7. Vertical dashed lines mark the approximate peak times of flares originating from AR 12561 (three B-class and one C-class). The gray shaded areas indicate the temporal coverage of the five observed scans.} \label{fig:goes}
\end{figure}

Solar flares are commonly classified based on their peak soft X-ray flux (0.1--0.8 nm) measured by the Geostationary Operational Environmental Satellite (GOES) system. The classification uses letters (A, B, C, M, X) followed by a number, indicating increasing flux. According to GOES data, AR 12561 was moderately active during its disk passage. On 2016 July 7, it produced the flare under investigation, which peaked at 07:49 UT and was classified as a C5.1 event. Figure~\ref{fig:goes} shows the GOES X-ray light curves for this day. Prior to the C5.1 flare, the AR produced a few minor B-class flares (B3.9, B4.4, B4.9) a few hours before. Following the C5.1 event, the region's activity significantly decreased. Our SST/TRIPPEL-SP observations, indicated by the gray shaded areas in Fig.~\ref{fig:goes}, fortunately covered the crucial phases of the C5.1 flare, including its onset, peak, and gradual decay, under good and stable atmospheric seeing conditions.

\begin{figure*}[t]
\centering
\includegraphics[width=\linewidth]{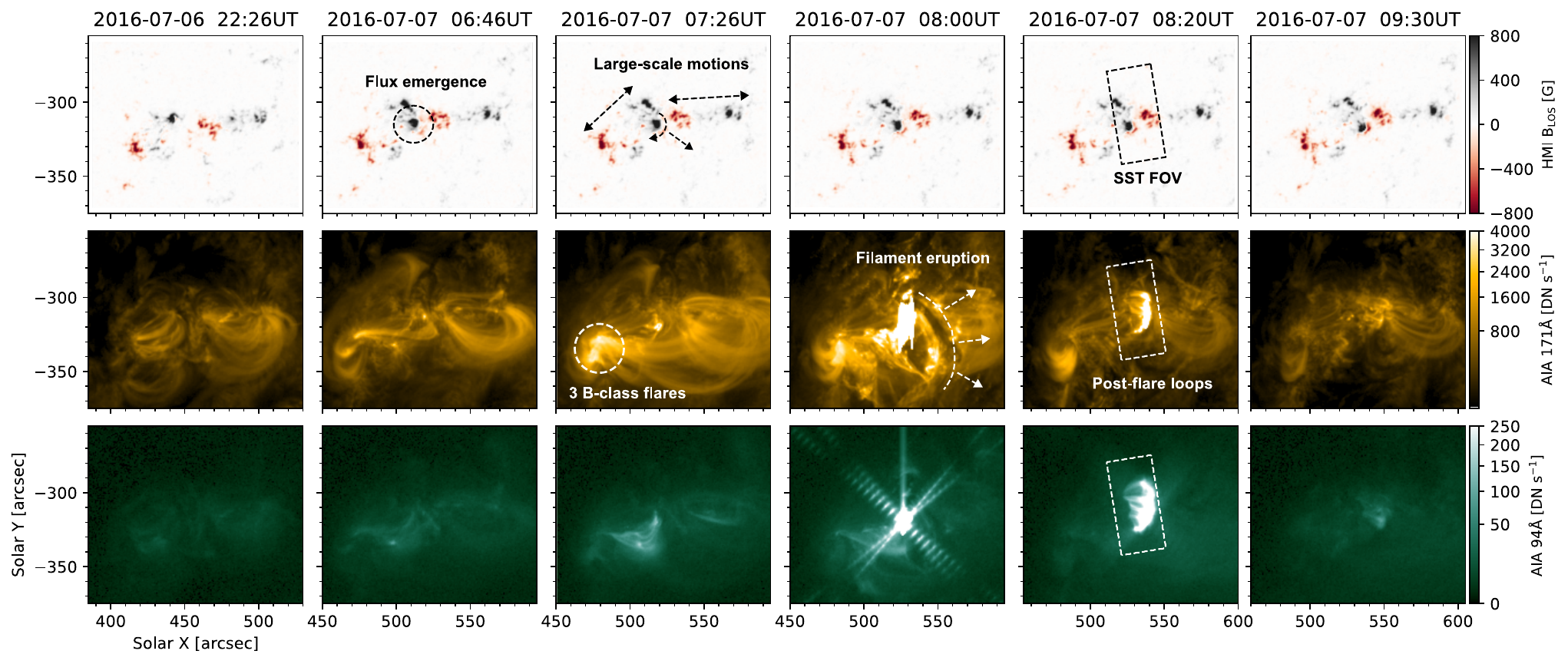}
\caption{Context images showing the evolution of AR 12561 on 2016 July 6-7. Panels show SDO/HMI line-of-sight magnetograms (top row), SDO/AIA 171\,\AA\ (middle row), and SDO/AIA 94\,\AA\ (bottom row) at four different times. The approximate FOV of the SST observations is indicated by the dashed white rectangle in the fifth panel.} \label{fig:mosaic}
\end{figure*}

\subsection{Large-scale context using SDO}\label{sec:AIA}

To place our high-resolution SST observations into a broader context, we utilized data from the Atmospheric Imaging Assembly \citep[AIA;][]{aia2012} and the Helioseismic and Magnetic Imager \citep[HMI;][]{hmi2012}, both instruments onboard NASA's Solar Dynamics Observatory \citep[SDO;][]{sdo2012}. AIA provides full-disk images in multiple EUV and UV wavelengths, probing different temperature regimes in the chromosphere, transition region, and corona, while HMI provides full-disk line-of-sight and vector magnetograms of the photosphere.

Figure~\ref{fig:mosaic} shows the evolution of AR 12561 from its emergence phase to the time of the flare. On July 6 (first column), HMI magnetograms reveal the emergence of at least two distinct magnetic bipoles in close proximity. The coronal structure seen in AIA 171\,\AA\ (sensitive to $\sim$1 MK plasma) shows relatively simple, separate loop systems associated with these bipoles. As the region evolved (second column, early July 7), further magnetic flux emerged near the center of the active region. This new flux included negative polarity which is canceled out by the pre-existing positive polarity, leaving a compact positive polarity concentration that developed into a prominent pore (visible as a strong magnetic concentration in HMI). The magnetic complexity of the region increased changing its connectivity, as evidenced by the intricate loop structures in the corona. The region exhibited minor flaring activity (B-class flares shown in Fig.~\ref{fig:goes}) originating from the eastern part of the region (third column of Fig.~\ref{fig:mosaic}).

After these episodes, continued shearing motions brought opposite polarities from the different emerging bipoles closer together, particularly at the interface region near the main positive pore. It is worth noting that the arrow tracing the pore’s motion in Fig.~\ref{fig:mosaic} is not straight, highlighting its rotational motion. This likely led to the build-up of magnetic stress and free energy. The AIA images show the presence of a dark filamentary structure overlying this polarity inversion line. This filament became unstable and started to erupt, coinciding with the onset of the C5.1 flare seen in GOES. The eruption is visible in the AIA channels (e.g., 171\,\AA) as rising cool material. Plane-of-sky velocity estimates from AIA image sequences suggest eruption speeds of approximately 70\kms. Following the filament eruption, the flare manifested as brightening in all AIA channels. As the flare progressed, the AIA channels (e.g., 171\,\AA\ and 94\,\AA) show the formation of a set of bright post-flare loops connecting the flare ribbons (fifth column). These ribbons correspond to the footpoints of the newly reconnected loops. The FOV of our SST/TRIPPEL-SP observations (dashed rectangle) was centered on this region, capturing the filament eruption and the subsequent flare development in high spatial and spectral resolution.

\begin{figure*}[t]
\centering
\includegraphics[width=0.98\linewidth]{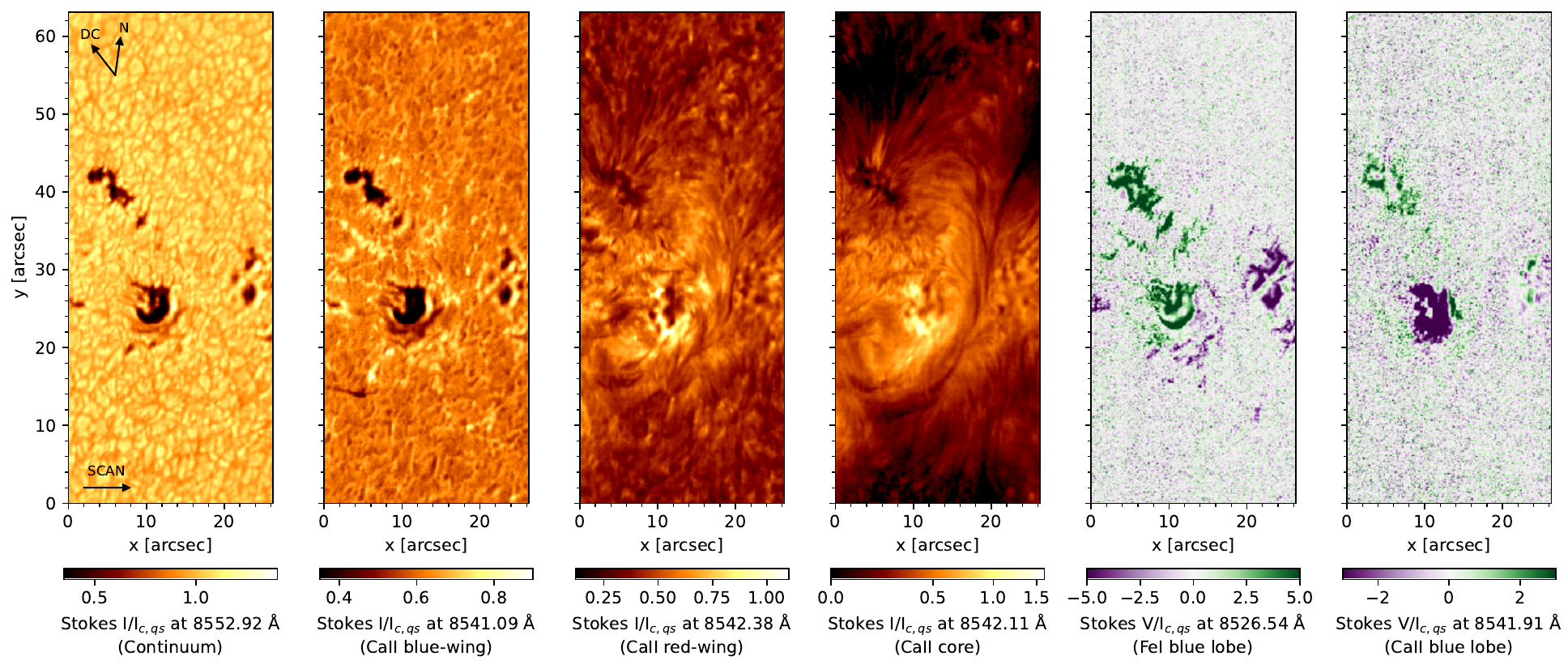}
\caption{High-resolution view of the target region from the first SST/TRIPPEL-SP scan (07:37-07:54 UT), taken just before the flare starts. From left to right: Intensity in the far wing of \ion{Ca}{ii} 8542\,\AA\ (photosphere); outer wing (upper photosphere); inner wing (lower chromosphere); and at the line core (mid-chromosphere). The last two panels show Stokes $V$ maps in the photospheric \ion{Fe}{i} 8526.67\,\AA\ line and the chromospheric \ion{Ca}{ii} 8542\,\AA\ line, respectively. The black arrows indicate the solar North and disk center. The horizontal axis corresponds to the scan direction (left to right). All panels are in units of the average quiet Sun continuum intensity.}
\label{fig:sstoverview}
\end{figure*}

\section{Methods}\label{sec:methods}

\subsection{Spectropolarimetric inversions}\label{sec:analysis}

We inferred the physical properties of the solar atmosphere by inverting the observed Stokes profiles using the STockholm Inversion Code\footnote{\url{https://github.com/jaimedelacruz/stic}} \citep[STiC;][]{delaCruz2016, delaCruz2019_STiC}. STiC is a massively parallel, multi-atom inversion code based on an optimized version of the Rybicki \& Hummer  \citep[RH;][]{Uitenbroek2001} radiative transfer code. In this work, we simultaneously inverted the chromospheric \ion{Ca}{ii} 8542\,\AA\ line and the photospheric \ion{Fe}{i} 8526.67\,\AA\ line. The \ion{Ca}{ii} atom was treated in NLTE using a 5-level plus continuum model. The \ion{Ca}{ii} 8542\,\AA\ line was computed assuming complete frequency redistribution (CRD). The photospheric \ion{Fe}{i} line was treated in LTE, since no photospheric signature of the flare was detected. Atomic data for \ion{Fe}{i} were taken from the Kurucz database\footnote{\url{http://kurucz.harvard.edu/}} and the Landé factors for the relevant energy levels were derived using J-K coupling. The spectral region also contains the photospheric \ion{Si}{i} doublet around 8556.8\,\AA, but preliminary tests showed difficulties in fitting it accurately, possibly due to atomic data inaccuracies, so it was excluded from the final inversions. The polarized radiative transfer equation was integrated using an efficient cubic DELO-Bezier formal solver \citep{delaCruz2013ApJ}. For each pixel (line-of-sight, LOS), STiC derives a 1D model atmosphere assuming a plane-parallel geometry. This approach is often referred to as 1.5D modeling, as it treats each pixel independently but provides a 3D view of the region.

STiC iteratively adjusts the physical parameters of the model atmosphere – specifically temperature ($T$), LOS velocity ($v_{\rm LOS}$), microturbulence ($v_{\rm mic}$), and the magnetic field vector ($\mathbf{B}$) – to find a synthetic spectrum that reproduces the observed profile. We note that the stratifications of gas pressure ($P_{\rm gas}$) and mass density ($\rho$) are calculated assuming hydrostatic equilibrium (HE), which may not be ideal for flaring regions. Nevertheless, this approximation has  been used by several authors \citep[e.g.,][]{Kuridze2017ApJ,Kuridze2018,Yadav2021_Cflare,Yadav2025_Ribbons,Kuckein2025A&A}, and remains reasonable outside the brief impulsive phase. Parameters are defined as functions of continuum optical depth at 5000\,\AA\ ($\log \tau$), and STiC modifies them at predefined node points along this scale. To avoid unphysical oscillations in the solution, the merit function includes a regularization term penalizing strong gradients or higher-order derivatives.

We initialized the inversion process using the FAL-C model atmosphere \citep{Fontenla1993}, interpolated onto a grid of 54 depth points spanning $\log \tau$ from $-7.0$ (transition region) to $+1.0$ (deep photosphere). We adopted an enhanced gas pressure at the upper boundary ($P_{\rm top} = 1.0$\dyn), representative of active region conditions \citep[cf.][]{delaCruz2019_STiC}. We initialized the magnetic field vector of the initial model atmosphere by using the result of applying the a neural-network-based implementation \citep{DiazBaso2025_neuralwfa}\footnote{\url{https://github.com/cdiazbas/neural_wfa}} of the spatially-constrained Weak-Field Approximation \citep{2020A&A...642A.210M} on the \ion{Fe}{i} and \ion{Ca}{ii} lines. We used the following node configuration: 9 nodes for temperature, 9 nodes for LOS velocity, 2 nodes for microturbulence, and 3 nodes for the LOS component of the magnetic field ($B_{\rm LOS}$). The transverse component of the magnetic field ($B_{\rm \perp}$) and its azimuth, initialized from the WFA, were kept fixed as constant values at the photosphere and chromosphere. Experimentation showed that adding nodes to these parameters, or increasing nodes for other quantities, did not significantly improve the overall fit quality.
%
%
%
Following an initial run on a reduced dataset (skipping every 4 pixels in both $x$ and $y$), a machine-learning approach was employed to generate initial guess models for the remaining pixels. This strategy helped mitigate divergence into local minima across neighboring pixels and eliminated non-converged solutions \citep{diaz_baso_bayesian_2022}.
\subsection{Non-force-free magnetic field extrapolations}\label{sec:nfff_extrapolations}

To investigate the 3D magnetic field structure and topology of AR 12561, we performed magnetic field extrapolations from photospheric vector magnetograms. While the solar corona is often approximated as force-free ($\mathbf{J} \times \mathbf{B} \approx 0$), the lower atmosphere (photosphere and chromosphere) is significantly non-force-free \citep[e.g.,][]{2001SoPh..203...71G}. Since the boundary data for extrapolations are photospheric measurements, using a non-force-free field (NFFF) model can, in principle, provide a more realistic representation of the magnetic field than other force-free approximations.

We employed the NFFF extrapolation technique developed by \citet{2008SoPh..247...87H} and \citet{2010JASTP..72..219H}, which is based on the principle of Minimum Dissipation Rate (MDR). This method assumes that the plasma-magnetic field system relaxes towards a state that minimizes the rate of energy dissipation. The resulting magnetic field can be represented as a superposition of a potential field ($\nabla \times \mathbf{B}_{\rm pot} = 0$) and two linear force-free fields ($\nabla \times \mathbf{B}_{i} = \alpha_i \mathbf{B}_{i}$, where $\alpha_i$ are constants), such that $\mathbf{B} = \mathbf{B}_{\rm pot} + \mathbf{B}_1 + \mathbf{B}_2$. The force-free parameters $\alpha_1$ and $\alpha_2$ are determined by minimizing the difference between the horizontal components of the extrapolated field ($\mathbf{B}_t$) and the inferred horizontal field ($\mathbf{b}_{t}$) at the photospheric boundary. The minimization target is the following metric $E_n$ \citep{Prasad2020ApJ}:
\begin{equation}
    E_n = \frac{\sum_{i=1}^{M} |\mathbf{B}_{t,i} - \mathbf{b}_{t,i}| \times |\mathbf{B}_{t,i}|}{\sum_{i=1}^{M} |\mathbf{B}_{t,i}|^2},
\end{equation}
where the sum is over all $M$ grid points on the photospheric boundary plane. This NFFF code\footnote{\url{https://github.com/AvijeetPrasad/MDR-NFFF}} has been successfully applied to study the magnetic configuration associated with various transient phenomena in active regions, such as Ellerman bombs, UV-bursts, flares, and coronal dimmings \citep[e.g.,][]{2018ApJ...860...96P,Prasad2020ApJ, 2023A&A...677A..43P, 2024A&A...691A.198J, Kumar2025SoPh}.

Due to the limited FOV, temporal cadence, and lower signal-to-noise ratio in the linear polarization signals of our TRIPPEL-SP observations, this dataset is not ideal for providing the boundary conditions for a reliable magnetic field extrapolation. Therefore, we used the magnetic field data from the SDO/HMI instrument, specifically the Spaceweather HMI Active Region Patch \citep[SHARP;][]{Bobra2014SoPh} data products. We used the hmi.sharp\_cea\_720s series, which provides vector magnetic field data in a Cylindrical Equal-Area (CEA) projection with a cadence of 12 minutes and a pixel scale of approximately 0.5\arcsec. We selected SHARP data for AR 12561 covering the time interval from 06:36 UT to 09:36 UT on 2016 July 7, encompassing the pre-flare, flare, and post-flare phases.

The computational domain for the NFFF extrapolations was a Cartesian box defined based on the selected SHARP patch. The horizontal extent covered 512 $\times$ 128 pixels, corresponding to approximately 256\arcsec $\times$ 64\arcsec. The vertical extent ($z$) was set to 128 pixels, corresponding to a height of approximately 64\arcsec above the photosphere. The NFFF code calculates the 3D magnetic field vector $\mathbf{B}(x, y, z)$ within this volume. From the extrapolated 3D magnetic field, we calculated magnetic field lines to visualize connectivity, the electric current density $\mathbf{J} = (c/4\pi) \nabla \times \mathbf{B}$ to identify regions of strong magnetic gradients and potential reconnection sites, and the magnetic free energy, computed as the difference between the total magnetic energy of the NFFF model and that of a potential field model derived from the same $B_z$ boundary condition. The free energy is given by:
\begin{equation}
    E_F = E_{\rm NFFF} - E_{\rm P} = \int_V \left( \frac{B^2_{\rm NFFF}}{8\pi} \right) dV - \int_V \left( \frac{B^2_{\rm P}}{8\pi} \right) dV,
\end{equation}
where $B_{\rm NFFF}$ and $B_{\rm P}$ are the magnetic fields from the NFFF and potential models, respectively, and $V$ is the extrapolation volume. We also estimated the column-integrated free energy density as:
\begin{equation}
    \Sigma_{\rm F}(x,y) = \Sigma_{\rm NFFF} - \Sigma_{\rm P} = \int_z \left( \frac{B^2_{\rm NFFF}}{8\pi} \right) dz - \int_z \left( \frac{B^2_{\rm P}}{8\pi} \right) dz,
\end{equation}
where $\Sigma_{\rm NFFF}(x,y)$ and $\Sigma_{\rm P}(x,y)$ are the column-integrated free energy densities for the NFFF and potential field models, respectively. This quantity provides insight into the spatial distribution of the magnetic free energy in the active region.

\begin{figure*}[t]
\centering
\includegraphics[width=\linewidth]{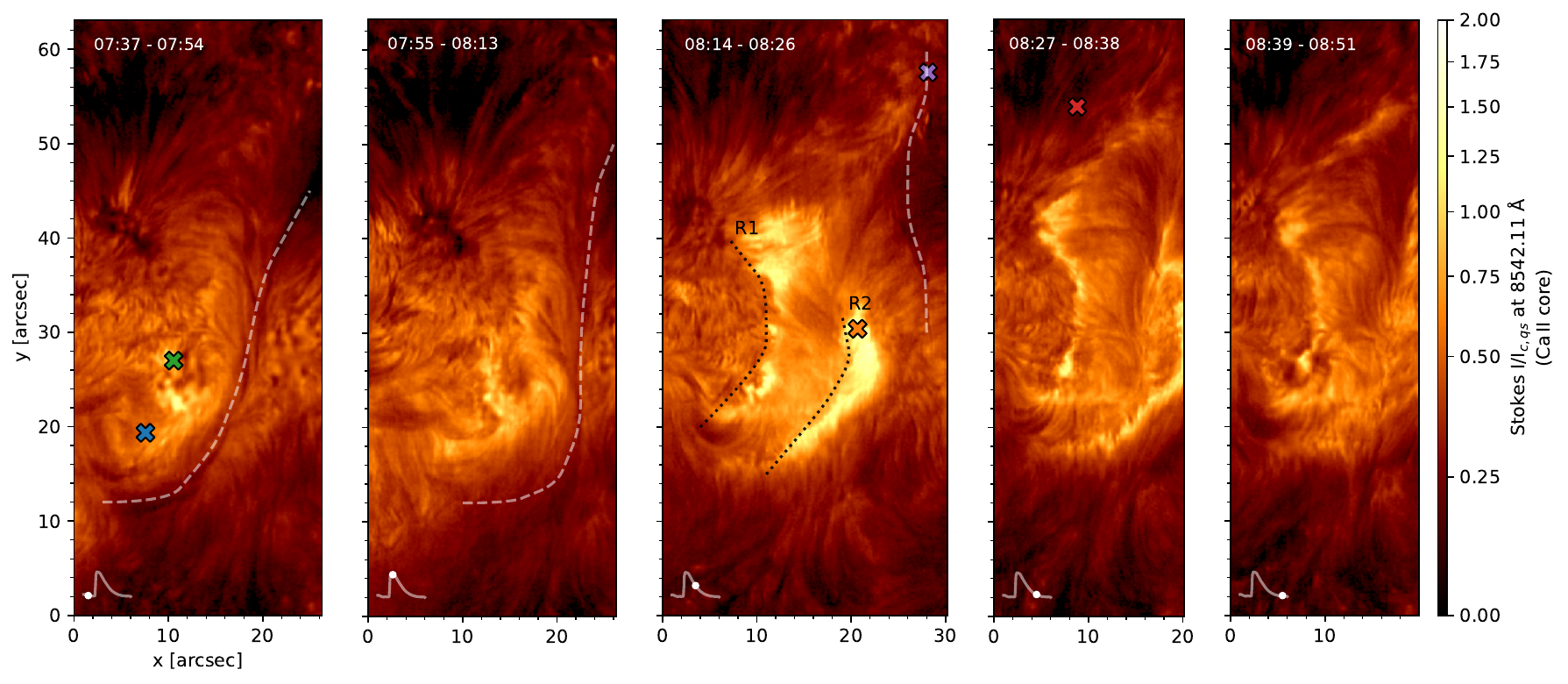}
\caption{Evolution of the active region as seen in the \ion{Ca}{ii} 8542\,\AA\ line core intensity,  from the five SST/TRIPPEL-SP scans. The time interval covered by each scan is indicated in the top left corner. The GOES 1-8\,\AA\ X-ray flux is plotted below each panel marking the mid-time of each scan. The location of the erupting filament's spine (inferred from multiple wavelengths) is indicated by the white dashed line and the flare ribbons (R1, R2) as dotted lines. Crosses indicate the location of the profiles shown in Fig.~\ref{fig:profiles}. 
} \label{fig:evolution}
\end{figure*}

\begin{figure}[t]
\centering
\includegraphics[width=\linewidth]{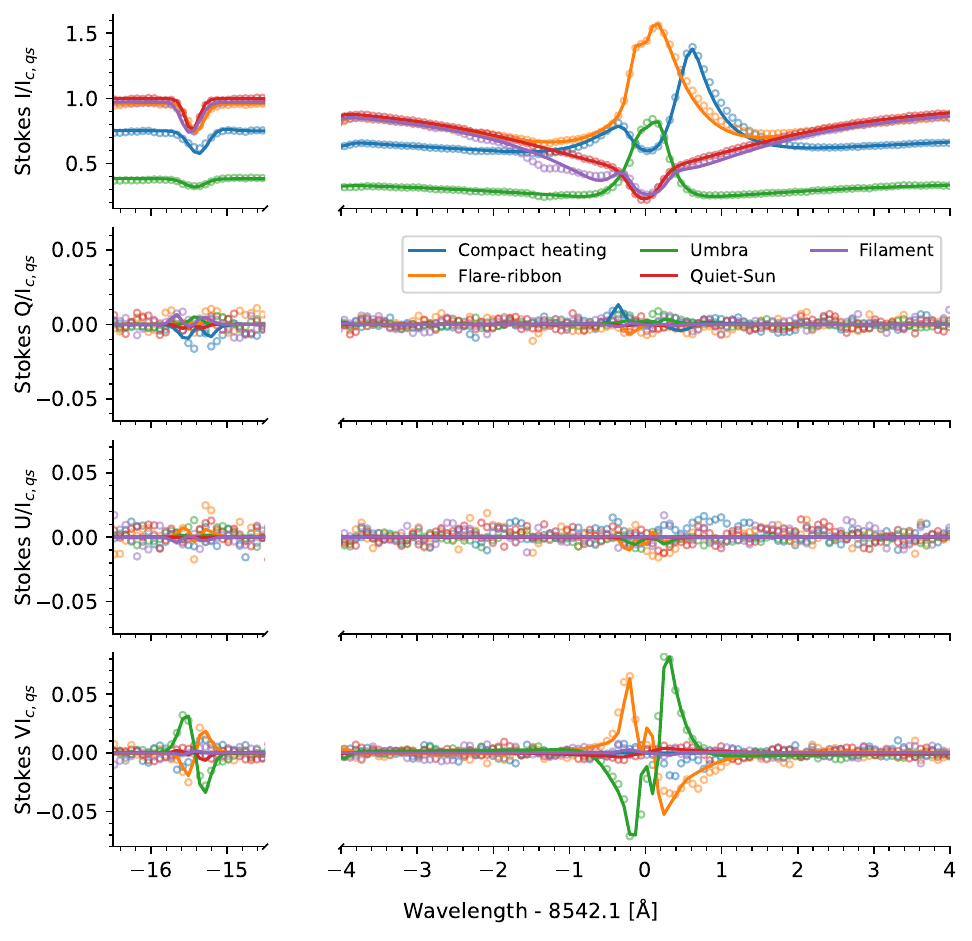}
\caption{Selection of profiles and their fits from different solar features. Their location is indicated in Fig.~\ref{fig:evolution}. It shows the Stokes profiles for the \ion{Fe}{i} 8526.67\,\AA\ line (left) and the \ion{Ca}{ii} 8542\,\AA\ line (right). Dots represent the observed profiles, while solid lines show the best-fit synthetic profiles from the NLTE inversions.}
\label{fig:profiles}
\end{figure}

\begin{figure*}[t]
\centering
\includegraphics[width=\linewidth, trim={0cm 1.1cm 0cm 0},clip]{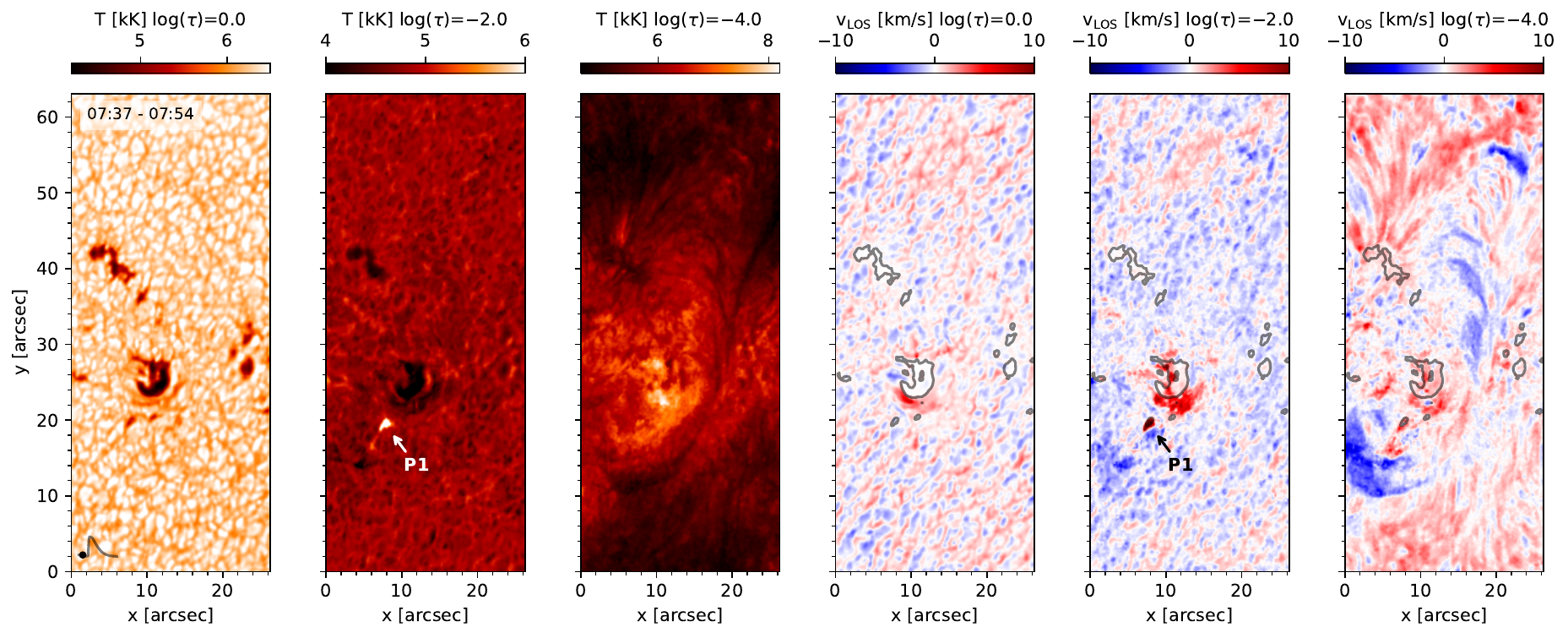} 
\includegraphics[width=\linewidth, trim={0cm 1.1cm 0 1.75cm},clip]{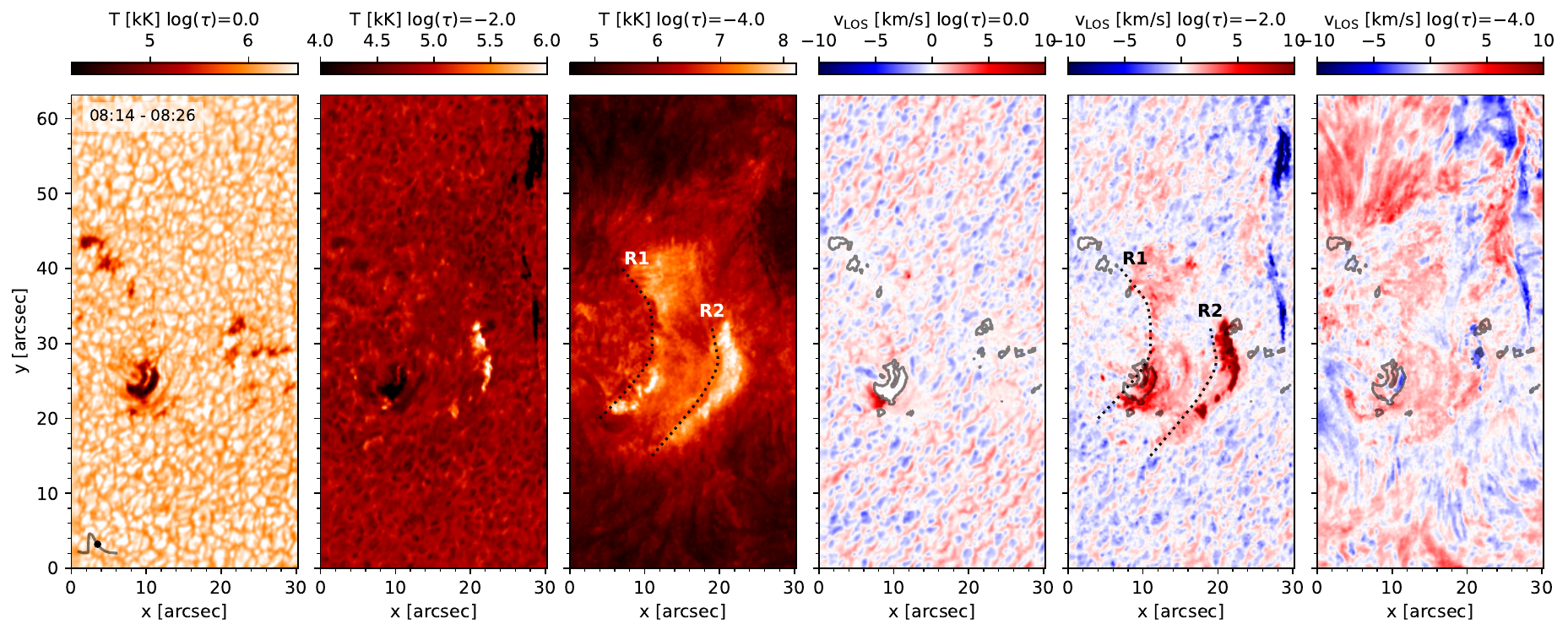} 
\includegraphics[width=\linewidth, trim={0cm 1.1cm 0 1.75cm},clip]{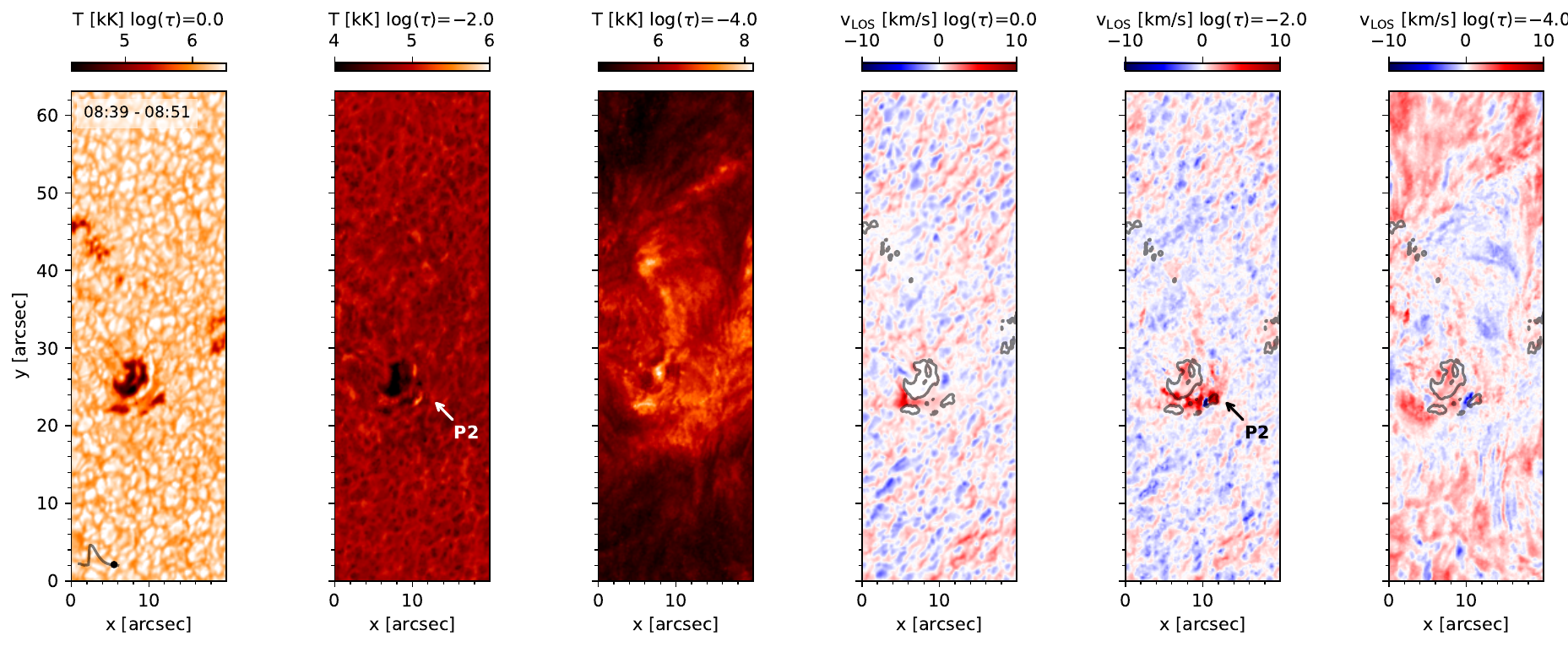} 
\caption{Atmospheric stratification inferred from NLTE inversions of \ion{Fe}{i} 8526\,\AA\ and \ion{Ca}{ii} 8542\,\AA\ for three representative scans: Scan~1 (07:36-07:48 UT, onset/pre-peak), Scan~3 (08:14-08:26 UT, post-peak/main phase), and Scan~5 (08:39-08:51 UT, decay). Each row corresponds to a scan. Columns show, from left to right: the temperature ($T$) and LOS velocity ($v_{\rm LOS}$) maps at three optical depths ($\log \tau = 0.0, -2.0, -4.0$). Positive (red) velocities indicate downflows, while negative (blue) velocities indicate upflows. Grey contours at $T=$5.5~kK ( $\log \tau = 0.0$) highlight the pores at the photosphere and dotted lines the location of the flare ribbons.}
\label{fig:inversions}
\end{figure*}

\begin{figure*}[t]
\sidecaption
\begin{minipage}{0.69\linewidth}
    \includegraphics[width=\linewidth]{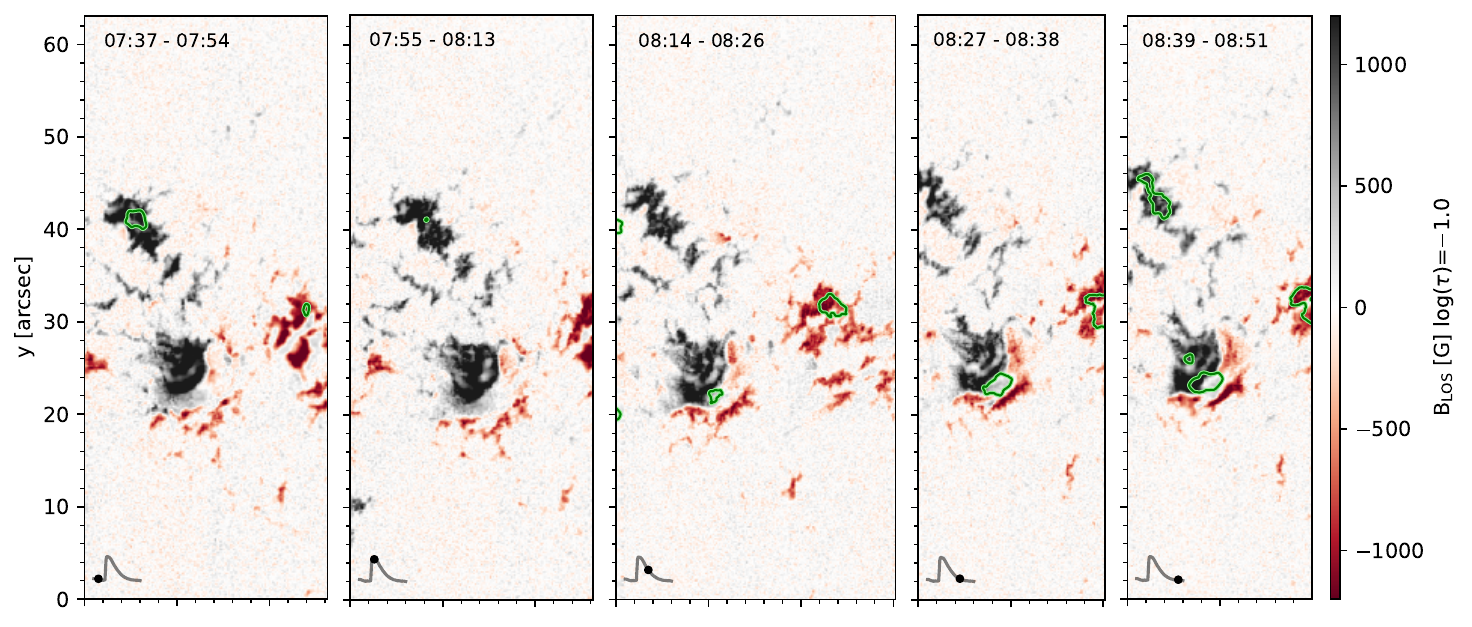}
    \includegraphics[width=0.9925\linewidth]{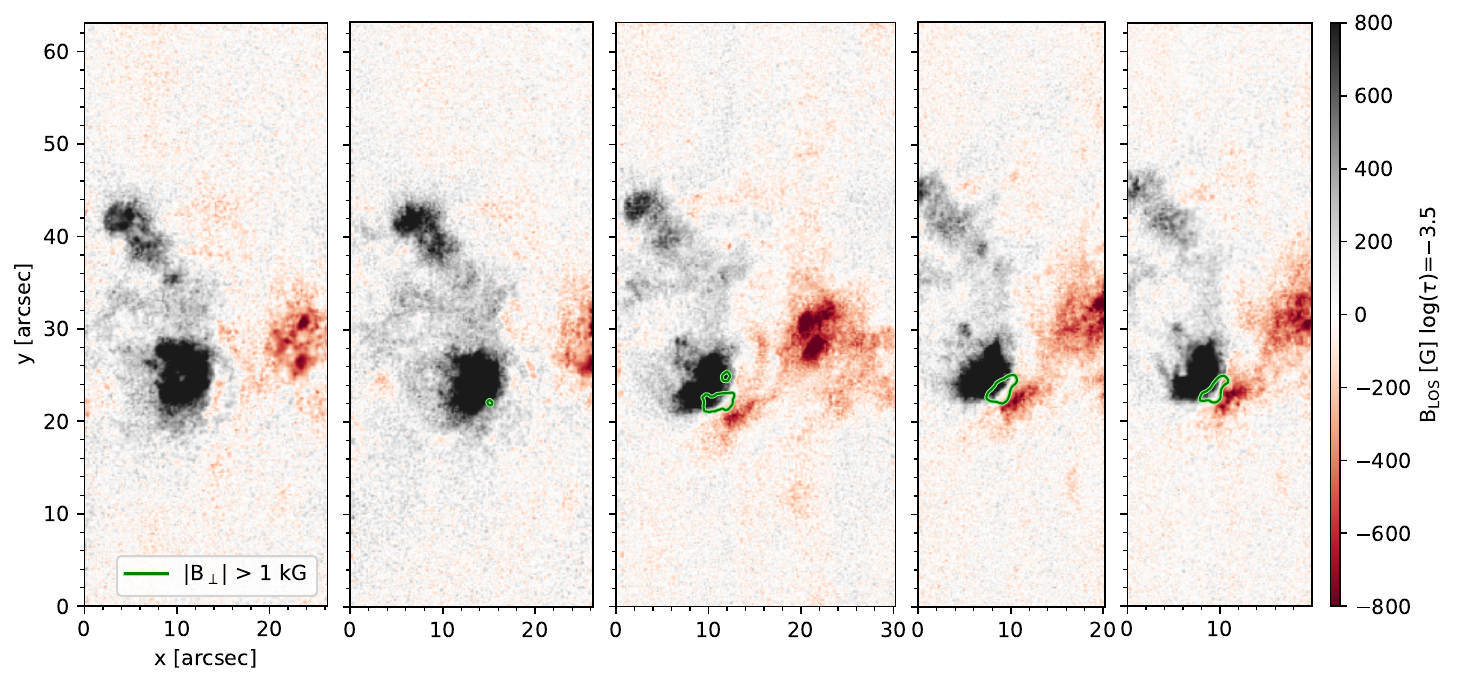}
\end{minipage}
\caption{Longitudinal magnetic field component ($B_{\rm LOS}$) at the photospheric level ($\log \tau = -1$) and in the chromosphere ($\log \tau = -3.5$) for all five scans, derived from NLTE inversions. Contours of the transverse component ($B_{\rm \perp}$) are overplotted for $B_{\rm \perp} > 1000$\,G for both maps. The time interval of each scan and the GOES 1-8\,\AA\ X-ray flux is plotted at each panel.}
\label{fig:magneticfield}
\end{figure*}

\section{Results}\label{sec:results}

\subsection{Observed dynamics at high spatial resolution}\label{sec:sstdata}

The high spatial resolution of the SST combined with the spectral information from TRIPPEL-SP allows for a detailed view of the photospheric and chromospheric dynamics before, during, and after the flare. Figure~\ref{fig:sstoverview} shows reconstructed intensity maps from the first scan (07:36-07:48 UT), just prior to the GOES flare peak, at different wavelengths across the \ion{Ca}{ii} 8542\,\AA\ line profile, illustrating the structures visible at different atmospheric heights.
The far wing image (leftmost panel) probes the photosphere, clearly showing granulation patterns and the prominent, dark pore near the center of the FOV (x$\approx$12\arcsec, y$\approx$25\arcsec). Surrounding the pore, particularly to the southwest, there is a diffuse region darker than the quiet sun but brighter than the umbra.
%
Moving closer to the line core (second panel), we observe the characteristic reversed granulation pattern, where intergranular lanes appear brighter than granule centers.
In the inner wing (third panel), the region above the pore appears relatively dark, but contains several small, localized bright patches. Fibril structures become more apparent across the FOV.
The line core image (fourth panel) is dominated by complex fibril patterns and, most notably, a long, dark filament structure snaking across the FOV from the northwest towards the southeast. This is the filament that subsequently erupts.

The last two panels of Fig.~\ref{fig:sstoverview} show a proxy for the LOS magnetic field: the circular polarization signals in the photospheric \ion{Fe}{i} 8526\,\AA\ line and the chromospheric \ion{Ca}{ii} 8542\,\AA\ line core. The \ion{Fe}{i} map confirms the strong positive polarity of the central pore and the surrounding negative polarity patches. The \ion{Ca}{ii} map shows significant magnetic signal in the chromosphere above the pore and in other active areas. Interestingly, the sign of the Stokes $V$ signal in the \ion{Ca}{ii} line above the pore appears opposite to that in the photosphere (\ion{Fe}{i} line). This sign reversal is likely due to a strong temperature gradient, which induce a change in the sign of the source function gradient with height. 

Figure~\ref{fig:evolution} presents a time sequence of the \ion{Ca}{ii} 8542\,\AA\ line core intensity maps reconstructed from the five TRIPPEL-SP scans, covering the flare evolution from onset to decay. Before the flare onset (Scan 1, 07:36-07:48 UT), the filament is clearly visible as a dark structure in the line core. The second scan (Scan~2, 07:55-08:07 UT) started shortly after the GOES peak, and the filament structure began to erupt. The motion of the filament is highlighted in different panels with a dashed line. During the evolution, the filament gets less visible in the core of the line, and better seen towards the blue wing, because of the global rising velocity of the filament. After the peak of the flare (Scan 3, 08:14-08:26 UT), the filament is reaching the edge of the TRIPPLE-SP FOV, and two bright flare ribbons are well-developed on opposite sides of the original filament path. The western ribbon (R2) is noticeably brighter and more structured than the eastern ribbon (R1). The two last scans (Scan 4, 08:27-08:38 UT and Scan 5, 08:39-08:51 UT) show the gradual decay of the flare ribbons. By the end of Scan 5, the intense flare emission has subsided, although the region still appears active compared to the quiet Sun. This sequence clearly captures the key morphological changes associated with the filament eruption and the flare ribbon formation.

To understand some of the spectral features of our observations, Fig.~\ref{fig:profiles} displays the Stokes profiles from several locations. For example, the profiles at the edge of the filament shows secondary absorption features indicating different blue-shifted components at the filament up to $-1.5$\,\AA, which suggests velocities up to 50\kms. In contrast, a profile from the flare ribbon shows a core in emission with a strong red asymmetry, indicative of strong downflows. Profiles from the umbra display a very low continuum intensity with a core in emission and strong circular polarization signals.

\subsection{Atmospheric structure from spectropolarimetric inversions}\label{sec:inversions_results}

Our analysis of the NLTE inversions, conducted using STiC, reveals the  atmospheric stratification during different flare phases. As a measure of the quality of the fits and the complexity of the profiles,  Fig.~\ref{fig:profiles} display the best-fit synthetic profiles for some of the most prominent solar features.

\subsubsection{Temperature and velocity stratifications}\label{sec:tempvlos}
Figure~\ref{fig:inversions} presents maps of inferred temperature, and LOS velocity at selected optical depths ($\log \tau = 0.0, -2.0, -4.0$) for three representative scans: Scan 1 (onset/pre-peak), Scan 3 (post-peak/main phase), and Scan 5 (decay phase). Intermediate scans (2 and 4) are provided in Appendix~\ref{sec:appendixA} (Fig.~\ref{fig:appendix}). We prioritize these selected optical depths as inversions become less constrained at greater altitudes ($\log \tau < -5.0$). Additionally, the inferred microturbulent velocity maps (not shown), exhibiting generally low values (2-5~\kms), showed no systematic variations with the flare's location \citep[e.g.,][]{Yadav2025_Ribbons} and are thus not discussed further in this section. 

The initial phase of the flare (Scan 1; 07:36-07:48 UT), characterized as the onset/pre-peak, reveals distinct atmospheric conditions. At the photospheric level ($\log \tau = 0.0$), the temperature map clearly shows the cool pore and surrounding granular structures. In the upper photosphere ($\log \tau = -2.0$), the average temperature is around 4800~K. Notably, a distinct, compact region of enhanced temperature (hereafter P1, located near x$\approx$8\arcsec, y$\approx$20\arcsec, adjacent to the pore) is visible, reaching temperatures $\sim$1000-2000~K higher than its immediate surroundings. This heating is strongly localized in height, being most prominent around $\log \tau = -2.0$, with little signature in the photosphere or the middle chromosphere ($\log \tau = -4.0$). Associated with this heating are strong downflows, peaking at 10-20\kms\ also around $\log \tau = -2.0$. Those high temperatures and velocities are indeed needed to explain a profile which has not only a strong emission in the core but a very strong asymmetry to the red, as displayed in Fig.~\ref{fig:profiles}. This localized, persistent heating and downflow signature, present also in Scan 2 (see Appendix Fig.~\ref{fig:appendix}), is a key finding. Given its location near the main pore and its occurrence deep in the atmosphere prior to the main flare, it  suggests precursor magnetic reconnection activity, related to the complex magnetic topology in this region and further investigated in Sect.~\ref{sec:extrapolation_results} \citep[cf.][]{DiazBaso2021, Leenaarts2025}. Elsewhere at $\log \tau = -2.0$, moderate downflows ($\sim$4\kms) are seen consistently over the pore itself. In the middle chromosphere ($\log \tau = -4.0$), the temperature map reflects the presence of the cool filament material (of about 5000\,K). The velocity map at this height shows coherent upflows associated with the filament, reaching about $-2$\kms\ along its spine and up to $-5$\kms\ near its footpoints, indicating the initial phase of its eruption.

Transitioning to the post-peak/main phase (Scan 3; 08:14-08:26 UT), our observations capture the atmosphere significantly after the filament's eruption, with flare ribbons fully developed. While the photospheric structure largely persists from the pre-flare state, the chromosphere exhibits dramatic transformations. Notably, the localized heating P1 observed during the first scan is no longer prominent. Instead, the two bright flare ribbons, clearly depicted in Fig.~\ref{fig:evolution}, emerge as the dominant features. Inversion results reveal substantial temperature enhancements within these ribbons. At $\log \tau = -4.0$, the eastern  ribbon (R1) reaches peak temperatures of approximately 6000~K, exceeding the western ribbon's  (R2) peak of $\sim$8500~K. This heating extends to deeper layers; at $\log \tau = -2.0$, R2 maintains a temperature increase of 1000-2000~K, while the R1 shows less heating at this depth, suggesting stronger or deeper energy deposition in the footpoint of R2. Concurrent with these temperature increases are strong chromospheric downflows, characteristic of chromospheric condensations driven by flare heating. Specifically, the ribbon R2 at $\log \tau = -2.0$ exhibits downflows up to 10\kms, whereas the R1 shows downflows around 5\kms. These downflows are primarily confined to these heights, with no discernible corresponding signature in the photosphere ($\log \tau = 0.0$) or higher layers. At the middle chromosphere ($\log \tau = -4.0$), the velocity field is more intricate, featuring patches of both upflows and downflows within and surrounding the ribbons, likely reflecting a complex interplay of ongoing evaporation and condensation dynamics. Some upflows appear near the edge of R2, but their reliability is limited, as they occur in the pixels with the strongest downflows and may result from overestimating velocities in regions with steep velocity gradients. Finally, this map also captures the last moments of the filament eruption, with upflows larger than 10\kms. Note, for example, that the subtle absorption features close to $-50$\kms\ in the filament profile shown in Fig.~\ref{fig:profiles} was not fully reproduced in the inversion.

As the flare progresses into its decay phase (Scan 5; 08:39-08:51 UT), the pronounced flare signatures observed previously notably diminish. Chromospheric temperatures within the former ribbon locations have decreased significantly compared to Scan 3, though they remain slightly elevated above pre-flare levels (e.g., average ribbon temperature is $\sim$7000~K at $\log \tau = -4.0$). The strong downflows observed in the ribbons during Scan 3 have also weakened considerably, with typical velocities returning to pre-flare values (typical values are < 2~\kms). Interestingly, a persistent downflow over the pore (termed P2), is visible, likely a signature of the still complex emerging magnetic flux interacting with the existing field. Overall, the thermal and dynamic structure of the chromosphere progressively relaxes towards its quiescent state during this phase.

\subsubsection{Magnetic field information from inversions}\label{sec:magneticfield_results}

The inversions provide the stratification of the magnetic field vector. Due to the low S/N in Stokes $Q$ and $U$, we focus on the LOS component of the magnetic field ($B_{\rm LOS}$) and only show contours of the transverse component ($B_{\rm \perp}$) at two representative heights: the photospheric level ($\log \tau = -1.0$) and a mid-chromospheric level ($\log \tau = -3.5$). Figure~\ref{fig:magneticfield} shows the $B_{\rm LOS}$ maps for all scans and the corresponding contours of $B_{\rm \perp} > 1000$\,G.

The photospheric $B_{\rm LOS}$ maps (top row of Fig.~\ref{fig:magneticfield}) consistently show the strong positive polarity of the pore and the surrounding negative polarity patches, which begin to compact around the pore. The chromospheric $B_{\rm LOS}$ maps (bottom row of Fig.~\ref{fig:magneticfield}) reveal a more dynamic behavior. While the overall pattern reflects the upward expansion of the photospheric field, notable changes occur, particularly near the main pore. Prior to and during the early phase of the flare (Scans 1 and 2), the chromospheric $B_{\rm LOS}$ above the pore is predominantly positive. However, following the main flare onset (Scan 3 and onwards), a clear intrusion of negative $B_{\rm LOS}$ polarity becomes apparent in the chromosphere immediately to the south/southwest of the main positive pore, reaching values of about $-$800\,G. Concurrently, the contours of strong transverse field ($B_{\rm \perp} > 1000$\,G) also exhibit changes. In the photosphere, there is a subtle enhancement of $B_{\rm \perp}$ regions near the pore, while in the chromosphere, the area covered by strong $B_{\rm \perp}$ near the pore appears to increase  from Scan 3 onwards. This chromospheric enhancement in $B_{\perp}$ and changes in $B_{\rm LOS}$, reflect the emergence of more horizontal magnetic fields into the chromosphere, potentially triggering the flare. 

These observed changes in both $B_{\rm LOS}$ and $B_{\rm \perp}$ provide direct evidence of magnetic-field evolution during the flare. Although the inversions of the observational data offer valuable insights into the chromospheric magnetic field dynamics, we turn to the NFFF extrapolations for a more comprehensive 3D view of the magnetic field evolution and its relation to energy release.

\begin{figure}[t]
\centering
\includegraphics[width=\linewidth]{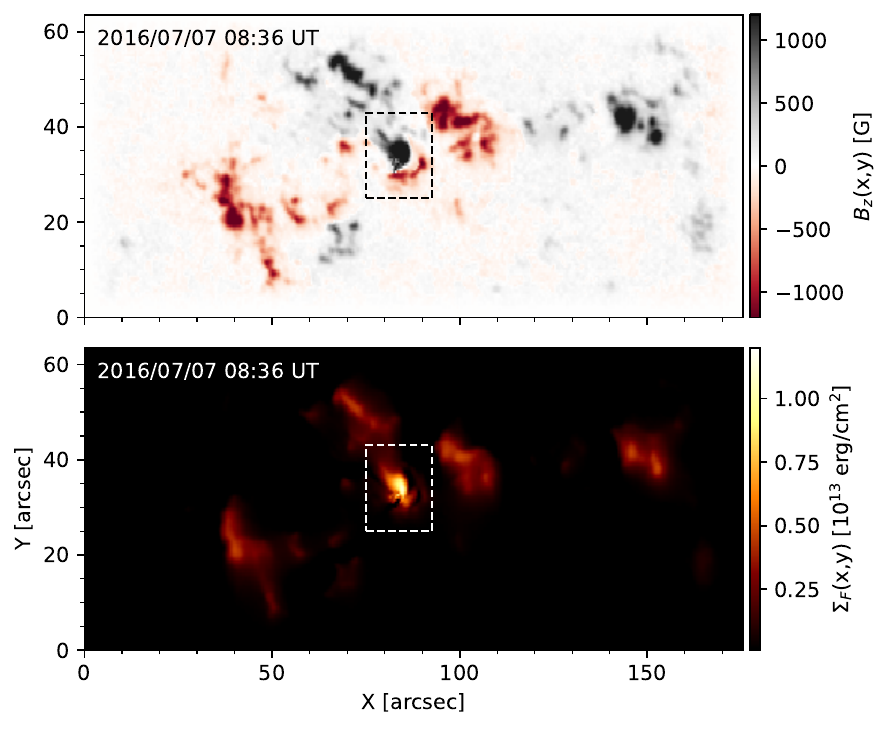}
\caption{Vertical magnetic field ($B_z$) at the photospheric level and column-integrated free energy density $\Sigma_F(x,y)$ at 08:36~UT, approximately at the peak of the flare. The $\Sigma_F(x,y)$ map highlights the spatial distribution of free magnetic energy, with the highest values concentrated at the main pore. The white dashed contour outlines the area used for total flux integration in Fig.~\ref{fig:free_energy}.}
\label{fig:free_energy_2D}
\end{figure}

\begin{figure}[t]
\centering
\includegraphics[width=\linewidth]{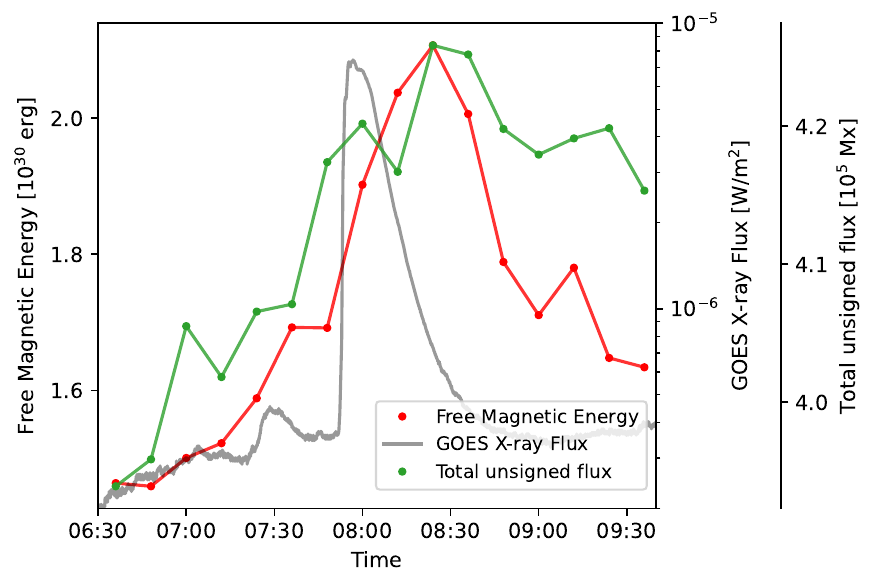}
\caption{Evolution of magnetic properties derived from NFFF extrapolations between 06:36 and 09:36 UT on 2016 July 7. It displays the GOES 1-8\,\AA\ X-ray flux, the total unsigned magnetic flux ($\Phi$) over the pore area (white dashed contour in Fig.~\ref{fig:free_energy_2D}), and the total free magnetic energy ($E_F$) within the same area.}
\label{fig:free_energy}
\end{figure}

\subsection{Magnetic field topology and free energy from extrapolations}\label{sec:extrapolation_results}

To understand the 3D magnetic context of the localized deep heating and the flare dynamics observed in the TRIPPEL-SP data, we analyzed the NFFF extrapolations performed using SDO/HMI vector magnetograms as boundary conditions.

\subsubsection{Magnetic Free Energy Evolution and Flare Energetics}

Our investigation into the flare's energetics begins by examining the spatial distribution of stored magnetic energy. Figure~\ref{fig:free_energy_2D} illustrates the vertical magnetic field ($B_z$) at the photosphere alongside the column-integrated free energy density $\Sigma_F(x,y)$ at 08:36 UT, coinciding with the C5.1 flare's peak. This map clearly highlights the concentration of free magnetic energy directly above the main pore, indicating that this region is the primary site for energy storage and release during the flare. To quantify this, we calculated the total free magnetic energy ($E_F$) and total unsigned magnetic flux ($\Phi = \int |B_z|,dA$) within the pore area, outlined by the white dashed contour in Fig.~\ref{fig:free_energy_2D}. This region was chosen as representative of the active region's energy reservoir and to provide a cleaner view of the relevant magnetic evolution of this location. Figure~\ref{fig:free_energy} subsequently presents the evolution of these quantities from 06:36 to 09:36 UT, contextualized by the GOES 1-8\,\AA\ X-ray flux.

Analyzing the temporal evolution, the free magnetic energy ($E_F$) within the pore area exhibits a gradual accumulation in the hours preceding the flare. It rises from approximately $1.4 \times 10^{30}$~ergs at 06:36 UT to a peak of $2.2 \times 10^{30}$ ergs at 08:36~UT. This sustained increase shows a build-up of free energy within the active region, likely driven by ongoing flux emergence and shearing motions observed in Fig.~\ref{fig:mosaic}. Following this peak, the free energy experiences a significant drop of about $\Delta E_F \approx 0.6 \times 10^{30}$ ergs, representing a decrease of approximately 30\% from its peak value. This magnitude of released magnetic free energy is consistent and sufficient to power a C5.1 class flare, considering the typical energy estimates for such events range from $10^{29}$ to $10^{30}$ ergs \citep[e.g.,][]{Su2014_FMEvsFP}. This provides a strong quantitative link between the observed magnetic field changes and the flare's energy output. Furthermore, the substantial relative decrease in free energy after the flare peak aligns with the general characteristic of eruptive flares, where a significant portion of stored magnetic energy is released \citep[e.g.,][]{2023RAA....23i5025W}. When the analysis is extended to the full active region, the calculated $\Delta E_F$ is slightly larger, reaching approximately $1.4 \times 10^{30}$ ergs, which accounts for about 15\% of the total free energy at the peak, still well within the energy budget for a C-class flare.

A notable observation is that the peak in free energy occurs slightly after the GOES X-ray peak. This temporal delay can be attributed to a continuous build-up of new free energy within the active region, likely fueled by ongoing magnetic flux emergence, as shown by the evolution of the total unsigned magnetic flux shown in Fig.~\ref{fig:free_energy}. This interpretation aligns with statistical studies showing that significant decreases in free energy are mostly observed in the strongest flares, while in weaker events, newly emerging flux can offset energy release \citep{Karimov2024}. Additionally, inherent limitations of the NFFF extrapolation method, which relies on photospheric boundary conditions provided at 12~min temporal cadence, may contribute to this observed delay, as it might not fully capture the rapid magnetic field reconfigurations occurring during the most dynamic phases of the flare.

\begin{figure}[t]
\centering
\includegraphics[width=0.80\linewidth]{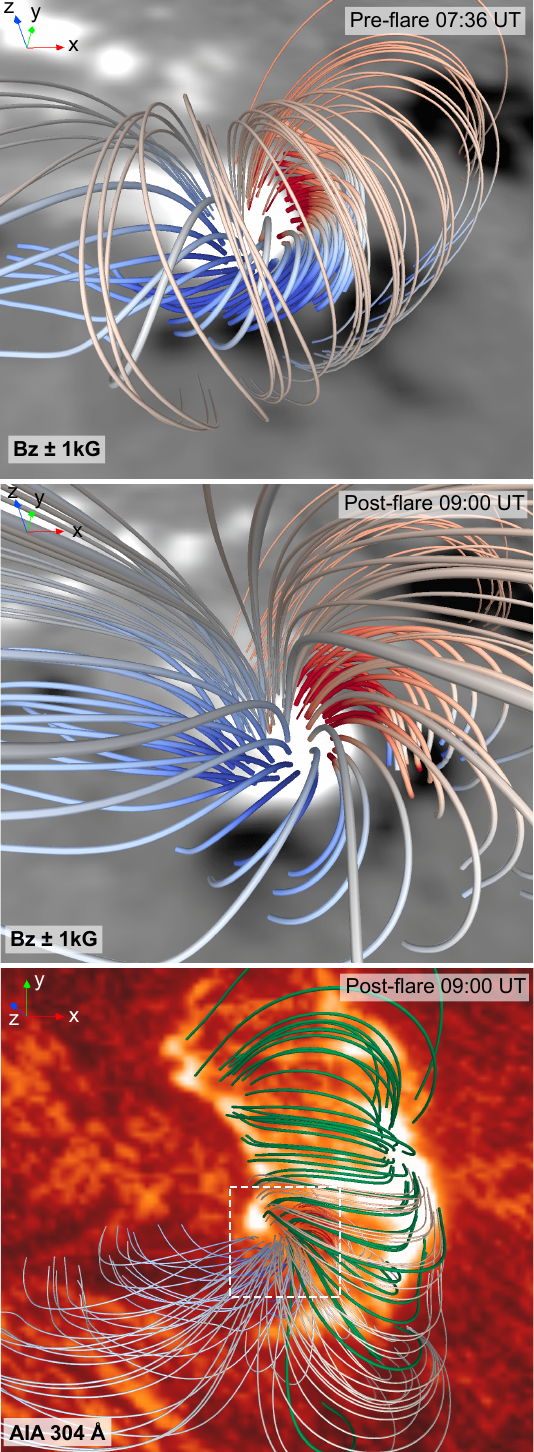}
\caption{Magnetic topology above the pore before (07:36 UT, top) and after (09:00 UT, middle and bottom) the flare. Field lines are colored by $B_x$ to highlight the shear. The bottom panel shows the post-flare arcade (green) overlaid on the AIA 304\,\AA\ image and the FOV of the previous two panels with a dashed-white square.}
\label{fig:topology_top}
\end{figure}

\begin{figure}[t]
\centering
\includegraphics[width=0.75\linewidth]{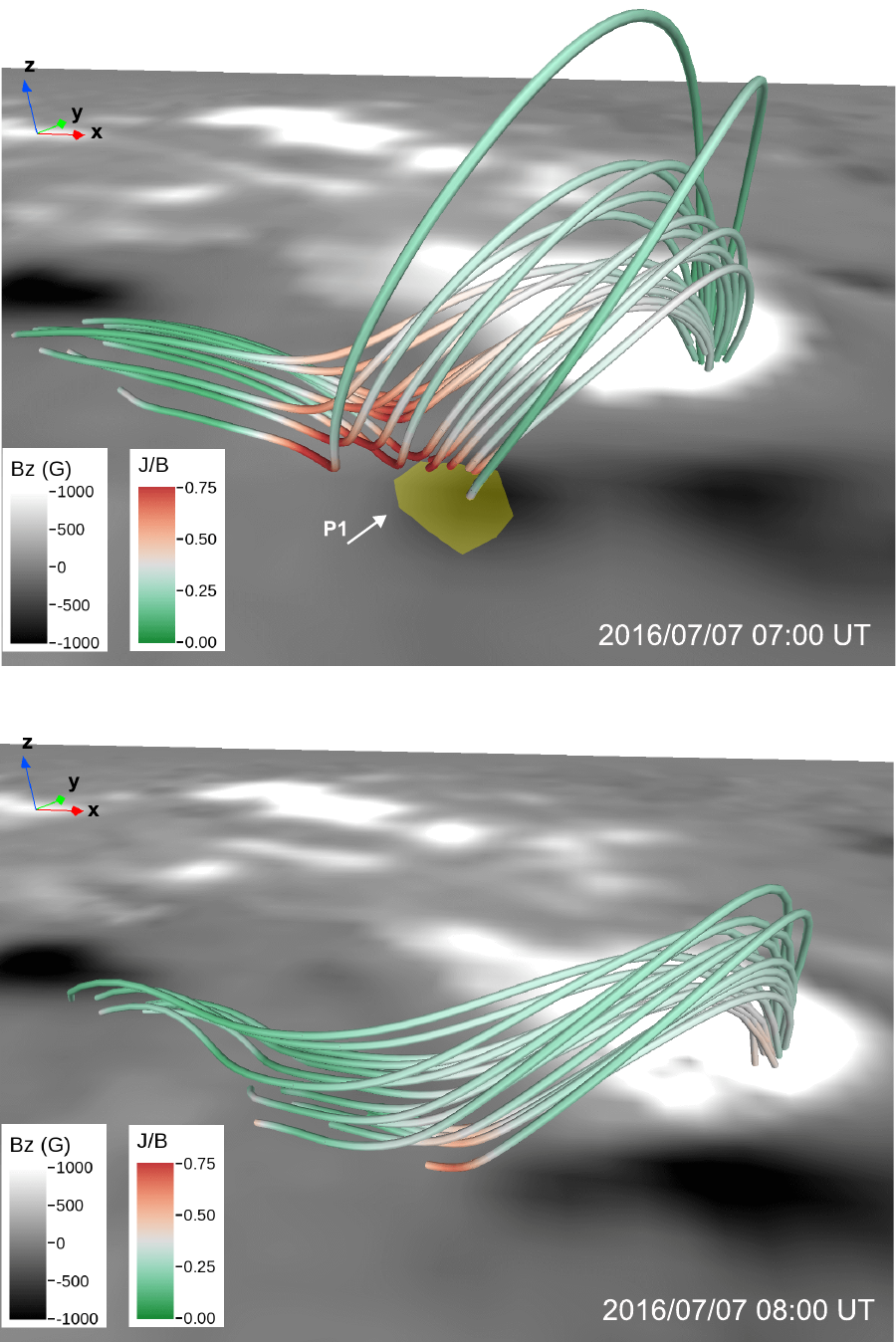}
\caption{Side view of the magnetic topology from NFFF extrapolations before the flare (top at 07:00~UT and bottom at 08:00~UT). The lower boundary shows photospheric HMI $B_z$ ($\pm$1000\,G). Field lines are colored by $J/B$ in units of pix$^{-1}$;  highlighting in red possible reconnection sites. The yellow area indicates the location of the localized heating P1 observed with TRIPPLE-SP.}
\label{fig:topology_baldpatch}
\end{figure}

\subsubsection{Magnetic Topology and Evolution}
The NFFF extrapolations offer a comprehensive illustration of the 3D magnetic field configuration above the pore and its dynamic evolution throughout the flare event. Key aspects of this topology are presented in Fig.~\ref{fig:topology_top} and Fig.~\ref{fig:topology_baldpatch}. The intricate magnetic field lines above the pore are displayed in Fig.~\ref{fig:topology_top} using the VAPOR visualization package \citep{vapor1}. Prior to the flare, specifically at 07:36 UT (top panel), the magnetic topology is dominated by highly sheared field lines. These structures, characterized by field lines running largely parallel to the polarity inversion line (PIL) rather than directly connecting opposite magnetic polarities, are a clear indicator of significant currents and stored magnetic energy, corroborating our free energy calculations. This magnetic configuration could have been facilitated by the flux emergence and rotational motion observed during 06:30-07:30~UT as shown earlier in Fig.~\ref{fig:mosaic}. Field lines are colored according to the $B_x$ component to visualize changes in the magnetic field's orientation along their path. In contrast, following the flare, at 09:00 UT, the magnetic field has undergone a drastic restructuring, with a radial magnetic field from the pore (center panel) and the formation of a classic post-flare arcade (highlighted in green, bottom panel). This new geometry shows a more relaxed and lower-energy state, which is a characteristic result of the release of magnetic energy. The contemporaneous AIA 304\,\AA\ image reveals bright emissions from the plasma within these post-flare loops, whose footpoints align with the flare ribbons observed by TRIPPEL-SP. Even though the extrapolations have captured the post-flare arcade, it did not capture the filament and its eruption, likely due to the weak magnetic fields at the filament footpoints hardly detectable in HMI.

For understanding the flare trigger, we investigated the magnetic topology close to the compact heating event termed P1. Figure~\ref{fig:topology_baldpatch} presents side views of the magnetic configuration of that region at two key times: 07:00~UT (pre-flare) and 08:00~UT (during the first TRIPPEL-SP scan). In these visualizations, field lines are colored by the $J/B$ ratio, effectively accentuating regions of high electric current density relative to the magnetic field strength. Strong red segments on field lines dipping down to the photosphere indicate the presence of bald-patch (BP) regions. These are defined as locations where the magnetic field $\mathbf{B}$ is tangent to the photosphere and exhibits an upward concavity. Such BP configurations are widely recognized as favorable sites for magnetic reconnection in the lower atmosphere \citep{2017RAA....17...93J, Aulanier2000}. The NFFF extrapolations indeed reveal the presence of these BPs near the main pore, particularly within areas characterized by high $J/B$. Significantly, these bald patches show a strong spatial correlation with P1, the site of localized deep heating and downflows observed prior to the main flare. This compelling co-location suggests that the observed activity is a direct signature of low-altitude magnetic reconnection occurring precisely at these bald patches. This specific reconnection event could therefore play an important role in channeling energy into the lower solar atmosphere and, importantly, in destabilizing the overlying magnetic system, including the erupting filament.

In summary, even though the NFFF extrapolations did not capture the large-scale filament and its eruption, they provide a coherent and comprehensive picture of the magnetic field's evolution throughout the flare event. This includes: the storage of energy within the highly sheared magnetic configuration above the pore, the identification of specific topological features that are prone to low-altitude reconnection and are co-spatial with the observed heating, and the subsequent dynamic transition to a relaxed post-flare arcade system. All these aspects are quantitatively consistent with the energy released during the flare.

\section{Discussion and conclusions}\label{sec:conclusions}

In this study, we have presented a multi-wavelength analysis of a C5.1-class solar flare and its associated filament eruption in AR NOAA 12561 on 2016 July 7. Our comprehensive approach combined high-resolution spectropolarimetric observations in the \ion{Ca}{ii} 8542\,\AA\ line and the \ion{Fe}{i} 8526\,\AA\ line from SST/TRIPPEL-SP, NLTE inversions using the STiC code, crucial context observations from SDO/AIA and HMI, and Non Force-Free Field magnetic field extrapolations based on HMI data. This combined analysis allowed us to construct a coherent narrative of the flare, tracing its evolution from the triggering mechanisms rooted in the pre-flare magnetic configuration to its profound energetic consequences in the chromosphere. Our key findings reveal several critical aspects of this flare event. 

Before the flare, a new flux emergence event was observed, which contributed to the complex magnetic configuration of the active region. This emergence was accompanied by significant shearing motions, particularly along the polarity inversion line where a long filament was present. The extrapolated magnetic fields revealed that most of the free magnetic energy was stored in this highly sheared magnetic field of the pore with sufficient free energy to power the observed C5.1 flare. During this pre-flare phase, we also identify a very compact localized heating event (termed P1) characterized by strong downflows and a temperature increase of 1000-2000 K, located near the main pore. This heating was spatially coincident with bald-patch regions identified in the NFFF extrapolations, suggesting that low-altitude magnetic reconnection occurred at these locations. 
The role of this activity in the flare's initiation is, however, open to interpretation. One scenario is that this reconnection directly contributed to the destabilization of the overlying filament, acting as a low-altitude trigger for the eruption \citep[e.g.,][]{Mitra2020}. An alternative view is that the primary driver was the persistent motion and flux emergence of the pore, which built stress and pushed the filament system towards  destabilization. In this latter scenario, the bald-patch reconnection would be a consequence of the evolution of the pore, rather than the primary trigger.

The filament eruption dynamics were clearly captured in both the \ion{Ca}{ii} 8542\,\AA\ line core intensity (Fig.~\ref{fig:evolution}) and its derived kinematics (Fig.~\ref{fig:inversions}, and Appendix Fig.~\ref{fig:appendix}). The NLTE inversions showed coherent upflows along the filament's structure in the middle chromosphere ($\log \tau = -4.0$). These flows initiated at approximately $-2$ to $-5$\kms\ during Scan 1 and accelerated to about $-10$\kms\ during the main eruption phase (Scan 2). Using the AIA images, we estimated the plane-of-sky velocity of the filament to be around 70~\kms, indicating that most of the filament's motion occurred perpendicular to the line of sight. While the NFFF extrapolations did not definitively resolve a distinct flux rope structure for the filament ---possibly due to resolution limits or the inherent complexity of the pre-eruptive field--- they nonetheless showed a highly sheared magnetic configuration along the polarity inversion line. The subsequent disruption of this sheared field and the associated decrease (of about 30\%) in magnetic free energy is energetically sufficient to power the observed C5.1 flare.

During the main and decay phases of the flare (Scans 3-5), the formation of two distinct flare ribbons was observed. The NLTE inversions quantified a notable chromospheric heating, reaching up to $\sim$6000 K in the eastern ribbon and $\sim$8500 K in the western ribbon at $\log \tau = -4.0$. Concurrently, strong downflows, indicative of chromospheric condensations, were detected, peaking at 5\kms\ in the eastern ribbon and 10\kms\ in the western ribbon at $\log \tau = -2.0$. These temperature enhancements and downflows represent characteristic signatures of energy deposition by non-thermal particles and/or thermal conduction along newly reconnected flare loops \citep[e.g.,][]{Kuridze2017ApJ, Yadav2021_Cflare}. At higher heights, one must be cautious interpreting the velocity maps: although some upflows were observed above the flare ribbons (and consistent with the interpretation of chromospheric evaporation), their reliability is limited. These upflows can be overestimated because of steep velocity gradients into the higher layers where the line is no longer sensitive, or simply the inversion code attempting to reproduce red asymmetries in the observed profiles as shown in \citet{Kuckein2025A&A}. This last point highlights the importance of using multiple spectral lines, especially in dynamic flare conditions.

Furthermore, our analysis revealed significant magnetic field evolution at chromospheric heights. Around the post-flare onset (Scan 3), the data inversions showed the appearance of a new patch of negative $B_{\rm LOS}$ ($\sim -800$\,G) near the pore, coupled with an increase of the chromospheric transverse field ($B_{\rm \perp}$) in this region (Fig.~\ref{fig:magneticfield}). However, it is not clear if the magnetic field has emerged to chromospheric heights as a result of magnetic reconfiguration or if the emergence of this new flux was the trigger of the filament eruption and flare. Magnetohydrodynamic simulations using the initial non-force-free field extrapolations as initial conditions could help to clarify this point, as has been done in previous studies \citep[e.g.,][]{2023A&A...677A..43P,Kumar2025SoPh}.

In conclusion, this work provides a detailed, self-consistent picture of the atmospheric response to a C-class flare and filament eruption, from trigger to decay. By linking high-resolution chromospheric spectropolarimetry from TRIPPEL-SP with NLTE inversions and 3D magnetic field extrapolations, we have highlighted the potential role of precursor reconnection at bald-patch locations in initiating the event, quantified the chromospheric impacts, and demonstrated the significant restructuring and energy release of the magnetic field. This reinforces the view that flare energy release is rooted in photospheric evolution, even if its most dynamic manifestations occur in the higher layers. Future studies combining such detailed chromospheric diagnostics with comprehensive coronal observations and more sophisticated magnetic field modeling will be crucial for a complete understanding of flare energy release and transport across the solar atmosphere.

\begin{acknowledgements}
CJDB acknowledge fruitful discussions with Prof. Lyndsay Fletcher and  Prof. Guillaume Aulanier.
%
This research is supported by the Research Council of Norway, project number 325491, and through its Centers of Excellence scheme, project number 262622, and Synergy Grant number 810218 459 (ERC-2018-SyG) of the European Research Council.
JdlCR gratefully acknowledges funding by the European Union through the European Research Council (ERC) under the Horizon Europe program (MAGHEAT, grant agreement 101088184).
The Swedish 1-m Solar Telescope is operated on the island of La Palma by the Institute for Solar Physics of Stockholm University in the Spanish Observatorio del Roque de los Muchachos of the Instituto de Astrof\'isica de Canarias. The Institute for Solar Physics is supported by a grant for research infrastructures of national importance from the Swedish Research Council (registration number 2021-00169). 
Data and images are courtesy of NASA/SDO and the HMI and AIA science teams.
We acknowledge the community effort devoted to the development of the following open-source packages that were used in this work: NumPy (\url{numpy.org}), Matplotlib (\url{matplotlib.org}), SciPy (\url{scipy.org}), Astropy (\url{astropy.org}), SunPy (\url{sunpy.org}), STiC, ISPy, and VAPOR (\url{vapor.ucar.edu}) for visualization.
This research has made use of NASA's Astrophysics Data System Bibliographic Services.
\end{acknowledgements}

\bibliographystyle{aa}
\bibliography{references}

\appendix

\section{SST/TRIPPEL-SP spectral range}\label{sec:fullrange}
\begin{figure}[h]
\centering
\includegraphics[width=\linewidth]{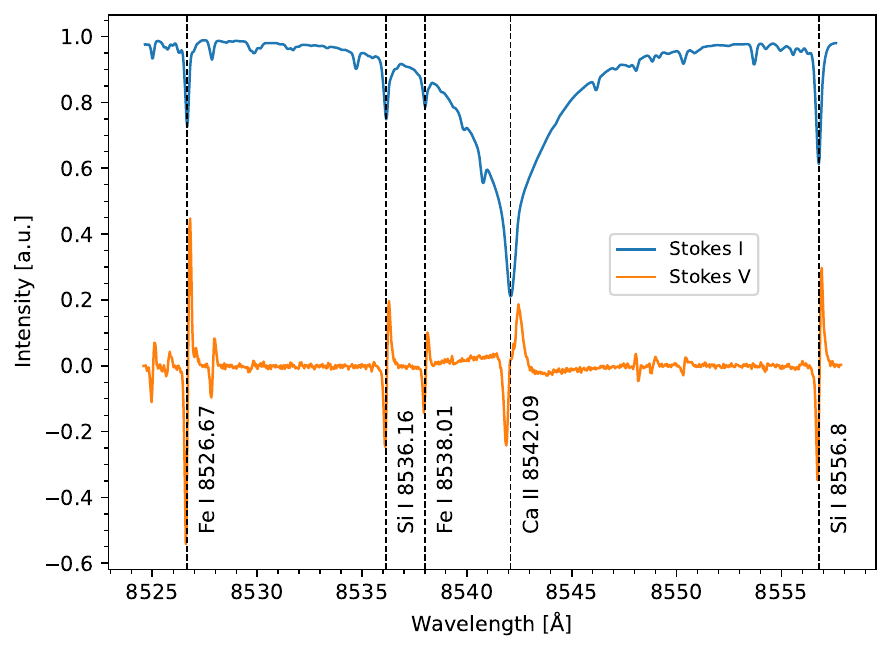}
\caption{Average quiet-Sun intensity spectrum and average Stokes $V$ profile from a magnetized region as a reference. Vertical lines highlight the main photospheric and chromospheric spectral lines of our spectral region.} \label{fig:im_qs_profile}
\end{figure}

The spectral range of our observational data is displayed in Fig.~\ref{fig:im_qs_profile} with the average intensity spectrum of the quiet Sun of our observations. We also include an average Stokes $V$ profile from the most magnetized region in our observations to show the relative sensitivity of each line to the magnetic field. Note that the noise level of individual pixels has a much larger amplitude, making the polarization signals coming from \ion{Si}{i} 8536\,\AA\ and the \ion{Fe}{i} 8538\,\AA\ difficult to detect.

\section{Additional inversion results for intermediate scans}\label{sec:appendixA}
\begin{figure*}[h]
\centering
\includegraphics[width=\linewidth, trim={0cm 1.1cm 0 0},clip]{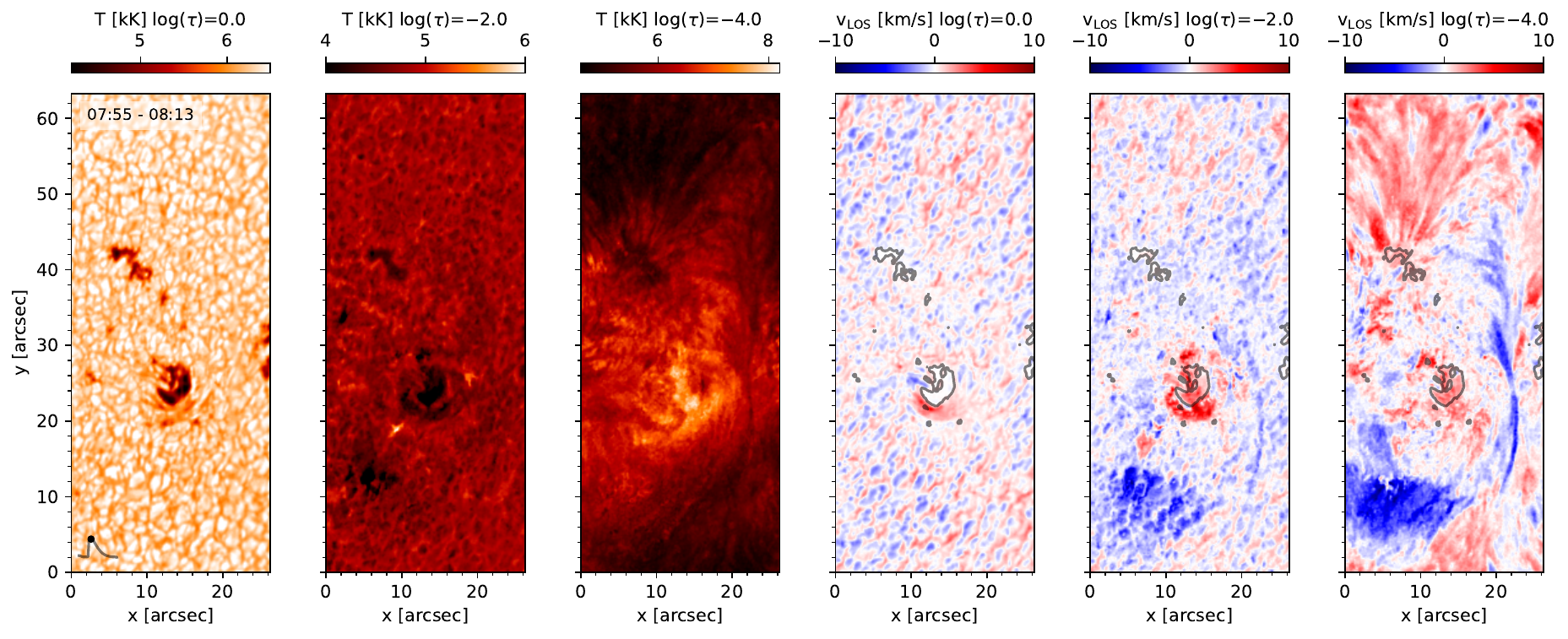} 
\includegraphics[width=\linewidth, trim={0cm 1.1cm 0 1.75cm}, clip]{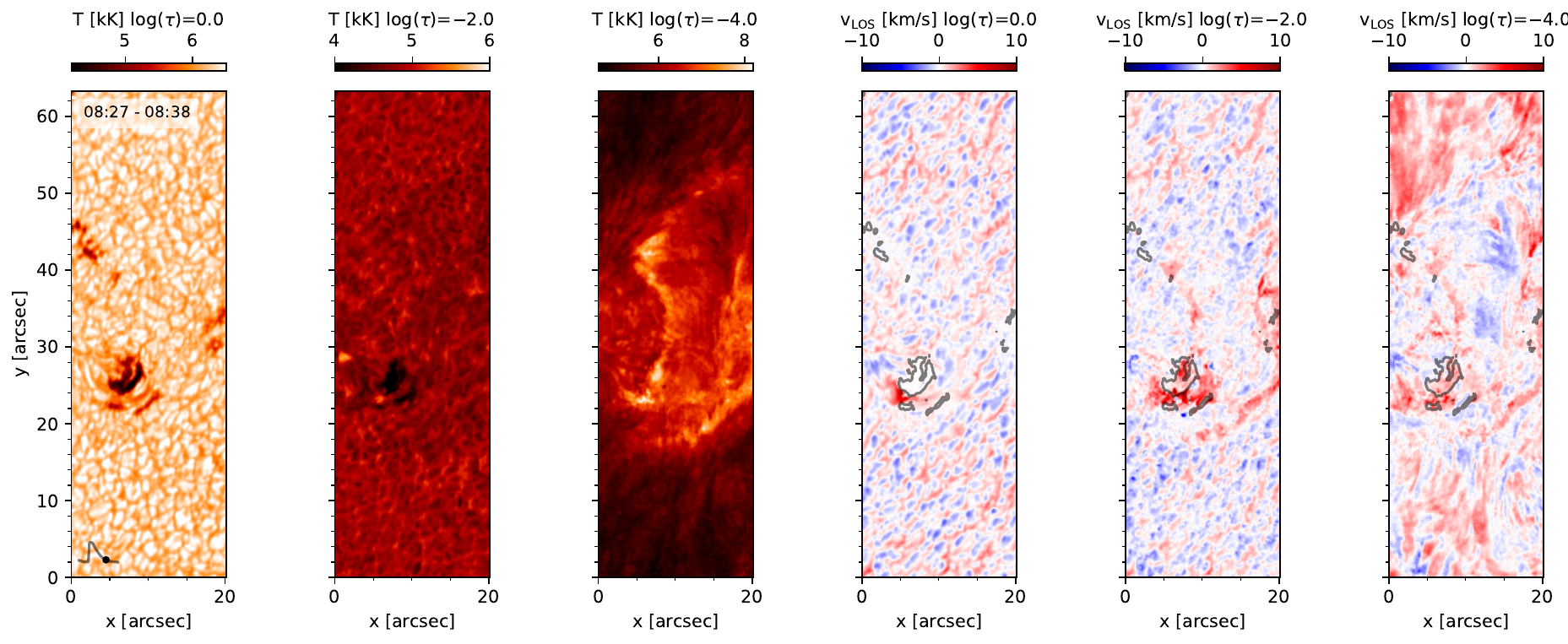} 
\caption{Atmospheric stratification from NLTE inversions for Scan 2 (07:55-08:07 UT, top row) and Scan 4 (08:27-08:38 UT, bottom row). Columns show, from left to right: temperature $T$ and LOS velocity at $\log \tau = 0.0$, $-2.0$, and $-4.0$.} 
\label{fig:appendix}
\end{figure*}

In this section, we present the inversion results for Scan 2 (07:55$-$08:07 UT) and Scan 4 (08:27$-$08:38 UT), which represent intermediate phases of the flare evolution. Figure~\ref{fig:appendix} shows the inferred atmospheric parameters in the same format as Fig.~\ref{fig:inversions}.

Scan 2 covers the period immediately following the GOES peak and captures the main phase of the filament eruption. The filament eruption is observed in the broad upflow signatures (reaching $\sim-10$~\kms) in the middle chromosphere ($\log \tau = -4.0$). Initial flare brightenings are visible, with moderate temperature ($\sim$8000~K) and associated downflows ($\sim$5~\kms) appearing in the chromosphere, before the fully developed ribbons seen in Scan 3. The localized heating in Region A is still present but starting to fade.

Scan 4 captures the transition from the main phase to the decay phase. The flare ribbons are still clearly visible but are less intense than in Scan 3. The chromospheric temperatures have decreased (e.g., eastern ribbon peak $\sim$7500~K at $\log \tau = -4.0$), and the associated downflows have weakened significantly ($\sim$2-4~\kms). The atmosphere is clearly relaxing towards its pre-flare state.

\end{document}